\documentclass[nofootinbib,amsmath,amssymb,showkeys,aps,prd,superscriptaddress]{revtex4-2}
\usepackage{booktabs}
\usepackage{lineno}
\usepackage{graphicx}
\usepackage[utf8,latin1]{inputenc}
\usepackage{soul}
\usepackage{mathrsfs}
\usepackage{subcaption}
\usepackage{caption}
\usepackage{academicons}
\usepackage{mathtools, nccmath}
\usepackage{tensor}
\usepackage{comment}
\usepackage[overload]{empheq}
\usepackage{tikz,xcolor}
\usepackage{cases}
\usepackage{setspace}
\usepackage{lipsum}
\usepackage[normalem]{ulem}
\usepackage{bm}
\usepackage{multirow}
\usepackage{soul}
\usepackage{hhline}
\usepackage{bigints}
\usepackage{hyperref}
\usepackage{amsmath}
\usepackage[T1]{fontenc}
\usepackage{fancyhdr}
\newcommand{\qm}[1]{``#1''}
\usepackage{hyperref}
\hypersetup{colorlinks=true, linkcolor={blue},citecolor={blue},urlcolor={blue}}  

\newcommand{\bsubeqs}{\begin{subequations}}
\newcommand{\esubeqs}{\end{subequations}}
\definecolor{lime}{HTML}{A6CE39}
\DeclareRobustCommand{\orcidicon}{
	\begin{tikzpicture}
	\draw[lime, fill=lime] (0,0) 
	circle [radius=0.16] 
	node[white] {{\fontfamily{qag}\selectfont \tiny ID}};
	\draw[white, fill=white] (-0.0625,0.095) 
	circle [radius=0.007];
	\end{tikzpicture}
	\hspace{-2mm}
}
\newcommand{\dd}{{\rm d}}

\newcommand{\ee}{{\rm e}}

\fancypagestyle{plain}{%
  \fancyhf{}
  \fancyfoot[C]{\iffloatpage{}{\thepage}}
  }
\foreach \x in {A, ..., Z}{%
	\expandafter\xdef\csname orcid\x\endcsname{\noexpand\href{https://orcid.org/\csname orcidauthor\x\endcsname}{\noexpand\orcidicon}}
}

\begin{document}
\doi{10.1103/ztky-t131}

\title{Families of regular spacetimes  and  energy conditions}

\author{Zi-Liang Wang\orcidA{}}
\email{ziliang.wang@just.edu.cn}
\affiliation{Department of Physics, School of Science, Jiangsu University of Science and Technology, Zhenjiang, 212003, China}

\author{Emmanuele Battista\orcidB{}}
\email{ebattista@lnf.infn.it}
\email{emmanuelebattista@gmail.com}
\affiliation{Istituto Nazionale di Fisica Nucleare, Laboratori Nazionali di Frascati, 00044 Frascati, Italy}

\begin{abstract}
We present a systematic method for constructing static, spherically symmetric regular spacetimes of Kerr-Schild type in general relativity that satisfy the weak energy condition. Our approach relies on physically reasonable assumptions on the matter  energy density, together with the boundedness of the Kretschmann scalar. The latter property ensures the finiteness of all curvature invariants and, for the configurations considered,  is equivalent to the completeness of causal geodesics. By classifying  admissible density profiles according to their complexity, we recover well-known regular black hole solutions such as the Bardeen, Hayward, and Dymnikova models, which are thus naturally embedded in a unified and broader framework.  Within this setting, we also derive closed-form expressions for several new families of regular geometries  involving hypergeometric  or incomplete Gamma functions, which in many cases  reduce to elementary functions including  algebraic, logarithmic, arctangent, and exponential forms. The emergence of horizons and photon spheres, as well as  their observational implications,  is also investigated. Finally, we extend the analysis to rotating metrics generated through the Newman-Janis algorithm.

\end{abstract}

\maketitle

%\tableofcontents

\section{Introduction}
\label{sec:intro}

Soon after Einstein formulated the final version of his famous field equations, the first exact solution describing a static,  spherically symmetric configuration was derived by Schwarzschild. This geometry exhibits two notable features:  the presence of an event horizon and a central point, $r=0$, where curvature becomes unboundedly large. Subsequent developments in our understanding of general relativity and its mathematical structure clarified that the  former corresponds to a coordinate singularity that can be removed via a suitable choice of coordinates, while the latter represents a true singularity, signaled by the incompleteness of causal geodesics. Similarly, homogeneous and isotropic cosmological scenarios, first devised by Friedmann shortly after Schwarzschild work,  indicate a universe  originated from a genuine singular state.

Singularities are generally  regarded as signaling a breakdown of Einstein theory. For this reason, the initial reaction of several authors (see e.g. Ref. \cite{Lifshitz1963}) was to interpret them as  artifacts of the high degree of symmetry underlying  the known exact solutions of Einstein equations \cite{Hawking-Ellis1973,Hawking1980,Wald}. Nevertheless, this viewpoint had to be withdrawn with the advent of Hawking-Penrose theorems, which proved that singularities do  arise as a generic consequence of the global causal structure of spacetime under physically reasonable assumptions \cite{Penrose:1964wq,Hawking:1970zqf,Senovilla1998,Mohajan2016}. These encompass the classical energy conditions, which encode standard  constraints on matter, such as the positivity of the energy density and the requirements underlying the focusing of geodesics  \cite{Curiel2017,Martin-Moruno2017}.

Despite their foundational importance, classical energy criteria are known to be not fulfilled in various semiclassical and quantum contexts, most notably in the Casimir effect and Hawking radiation \cite{Curiel2017,Martin-Moruno2017}, and even  at the classical level, with  traversable wormholes providing a prominent example \cite{Morris-Thorne1988,visser1995lorentzian}. These circumstances have motivated a more refined investigation of the interplay between energy conditions and the global structure of spacetime.

One particularly active area where this relationship becomes manifest is the construction of regular, singularity-free geometries, which are widely explored  in the literature, mostly as alternatives to the standard black hole paradigm \cite{Bambi2023-book,Lan:2023cvz,Myszkowski2025,Carballo-Rubio:2025fnc}. Following Bardeen seminal proposal,    based on an effective metric \emph{ansatz} smoothly  interpolating  between an asymptotically Schwarzschild-like exterior and a de Sitter-like interior \cite{bardeen1968non}, notable  de Sitter-core models were devised  by  Dymnikova \cite{Dymnikova1992},   Hayward \cite{Hayward2005}, and  Frolov and  Zelnikov \cite{Frolov2017}, along with numerous  subsequent extensions (a non-exhaustive list includes Refs.  \cite{Liu2026a,Singh2023,Zeng2025,Vertogradov2025b,Konoplya2024,Alshammari2025,Vertogradov2025,Fathi2021,Gohain2024,Bora2025,Waseem2025,Liang2025,Naseer2025}). In a closely related direction, regular Schwarzschild interiors smoothly joined to an exterior Schwarzschild region have also been recently investigated in Refs.~\cite{Ovalle:2025pue,Ovalle:2026lxb}, where infinite families of regular interiors were constructed by imposing regularity at the origin together with differentiability conditions at the Schwarzschild horizon. A different, field-theoretic interpretation emerged from Refs.~\cite{Ayon-Beato:1999kuh,Ayon-Beato:2000mjt}, where it was shown  that Bardeen metric can arise as  an exact  solution of Einstein gravity coupled to nonlinear electrodynamics, a framework that has since been extensively developed owing to its ability to  embed regular configurations into a consistent field-theoretic setting \cite{Dymnikova:2004zc,Balart2014,Culetu:2014lca,Fan:2016hvf,Singh:2022xgi,Bronnikov2022,AraujoFilho:2026sna}.  Other approaches comprise those incorporating quantum corrections \cite{Liu:2020ola,Cadoni2022,Sidharth2026,Borissova:2026dlz},  those sourced by a variable equation-of-state fluid \cite{Vertogradov2024c,Heidari2024}, and those bearing a Minkowski core \cite{Simpson:2019mud,Simpson:2021dyo}, as well as  interpolating black-bounce scenarios \cite{Simpson:2018tsi,Lobo2020f,Ditta2024a}, models  arising in extended theories of gravity \cite{Hu:2023iuw,Estrada2024,Bueno2024},  others exhibiting  
metric signature changes \cite{Capozziello:2024ucm,Zhang2026}, and primordial-black-hole dark matter scenarios \cite{Calza2024a,Calza2024b,Calza2025a}.

In many constructions, the regularization mechanism entails the violation of  one or more classical energy criteria within a finite domain,  consistently with  the implications of singularity theorems (see e.g. Ref. \cite{Borissova:2025hmj}). However, the precise connection between regularity and energy conditions is  subtle. Spacetime regularity imposes restrictions on the behavior of the metric  and curvature invariants, which in turn translate into relations for  the energy density and pressures through the Einstein equations. Whether these constraints are compatible with the energy criteria  depends sensitively on the functional form of the matter distribution. Therefore, a systematic and model-independent analysis in this direction is  desirable.

Motivated by these considerations, in this paper, we introduce a method for constructing  families of static, spherically symmetric  geometries of Kerr-Schild type that are regular and  satisfy at least the weak energy condition.  The primary input of our procedure is the matter energy density, which is subject to physically reasonable boundary and monotonicity assumptions. The ensuing solution of the Einstein field equations has  finite curvature invariants, a feature that also guarantees, under the hypotheses employed  in this work,  the completeness of causal geodesics. 

Within  our approach, several well-known regular models can  be understood as  belonging to a broader and unified framework,  in which they are recovered for particular choices of the parameters characterizing the energy-density profiles. In this general setting, enriched by the possibility of resorting to the superposition principle, new regular configurations  can  be naturally identified, and their compatibility with the energy conditions can be determined in a transparent manner.  This allows us to establish a complete pattern  connecting regularity, energy conditions, and the functional freedom underlying the construction of static, spherically symmetric metrics.  

The plan of the paper is as follows. In Sec.~\ref{sec:2}, we introduce the geometric setup, discuss curvature invariants,  and derive constraints ensuring regularity and compatibility with  the weak energy condition. Then, in Sec.~\ref{sec:3}, we construct and classify several families of regular solutions generated by different classes of energy-density profiles, organized according to their complexity.  Horizons and photon spheres associated with the resulting spacetimes, including  some implications for black-hole shadow phenomenology, are investigated in Sec.~\ref{sec:4}. Subsequently,  in Sec.~\ref{Sec:NJ-rotating}, we employ the Newman-Janis algorithm to build the rotating counterparts of the  regular models obtained via our scheme. Finally, we draw our conclusions in Sec.~\ref{sec:conclusions}. Supplementary material is given in the appendices.

\emph{Conventions.} We use   $G = c = 1$ units, and  metric signature $(-+++)$.

\section{General framework and regularity conditions}
\label{sec:2}

In this paper, we focus on  static,  spherically symmetric  metrics of  Kerr-Schild type \cite{visser1995lorentzian,Jacobson2007,Ovalle2023,Casadio2024} 
\begin{align}\label{eq:metric_static}
{\rm d}s^2
= g_{\mu \nu}{\rm d}x^\mu{\rm d}x^\nu
= - B(r)\,{\rm d}t^2
+ B^{-1}(r)\,{\rm d}r^2
+ r^2\,{\rm d}\Omega^2\,,
\end{align}
where ${\rm d}\Omega^2 = {\rm d}\theta^2 + \sin^2\theta\,{\rm d}\phi^2$ denotes the line element on the unit two-sphere. 

Typically, two distinct methods can be pursued to assess whether a solution is  regular \cite{Lan:2023cvz,Hu:2023iuw}. One relies on  the finiteness of  curvature scalars, while the other is a \qm{coordinate-based} approach  (in the sense that it involves, in practice, the choice of a coordinate system)  rooted in the completeness of causal geodesics. The two paradigms are generally not tantamount, as there exist models complying with  one notion while violating the other \cite{Geroch1968,Wald,Olmo2015}.  However, for the one-function  setup \eqref{eq:metric_static},  the two definitions were shown in Ref.~\cite{Hu:2023iuw} to be  equivalent for strictly asymptotically flat geometries possessing regular cores.  As we discuss below, this equivalence can be extended to a broader class of metrics displaying the (weaker) asymptotic Minkowskian behavior. Therefore, for such configurations, we can safely characterize regularity  by the absence of curvature singularities. 

Curvature invariants of the metric \eqref{eq:metric_static}  are examined in Sec. \ref{Sec:curvature-invariants}, where we also discuss  assumptions that guarantee their boundedness  both at an arbitrary  point $r=r_0>0$ and at the center $r=0$. Afterward,  in Sec. \ref{Sec:role-ECS},  we consider classical energy conditions and show that the weak and strong ones cannot be valid simultaneously. For this reason,  in Sec. \ref{Sec:sufficient-conditions}, we restrict our attention to the null and weak energy criteria, and develop a general procedure for constructing  regular Kerr-Schild spacetimes adhering to  them. 

\subsection{Geometry and curvature invariants}\label{Sec:curvature-invariants}

In the  geometry \eqref{eq:metric_static}, the Ricci, Kretschmann, and Weyl  scalars take the form, respectively,
\begin{subequations}\label{eq:curvaturescalar}
\begin{align}
R &= g^{\mu\nu}R_{\mu\nu}
= - B''(r)
- \frac{2\,[2 r B'(r)+B(r)-1]}{r^2}\,,\label{eq:R_scalar} \\
K &:= R_{\mu\nu\rho\sigma}R^{\mu\nu\rho\sigma}
= B''(r)^2 + \frac{4 B'(r)^2}{r^2}
+ \frac{4\,[B(r)-1]^2}{r^4}\,, \label{eq:K_scalar} \\
\mathcal{I}_1 &:= C_{\mu\nu\rho\sigma}C^{\mu\nu\rho\sigma}
= \frac{\left[r^2 B''(r)-2 r B'(r)+2 B(r)-2\right]^2}{3 r^4}\,,
\label{eq:I_scalar}
\end{align}
\end{subequations}
where  a prime denotes henceforth differentiation with respect to the radial variable $r$. 
In spherically symmetric spacetimes, some invariants involving the Hodge dual vanish identically, including  the Chern-Pontryagin scalar $K_2:= 
{}^*R_{\alpha\beta \mu\nu}R^{\alpha\beta \mu\nu}$ and the second Weyl invariant $\mathcal{I}_2 := {}^*C_{\alpha\beta \mu\nu} C^{\alpha\beta \mu\nu}$ ~\cite{Owen:2021eez} (which turn out to be equivalent \cite{Cherubini2002}). Other scalars can be expressed in terms of those listed above, such as  the Ricci  tensor squared and Euler invariant, which  read, respectively \cite{Obukhov1995,Cherubini2002,Steane2021,Jackiw:2003pm},
\begin{align}
R_{\mu\nu}R^{\mu\nu}
&= \frac{K-\mathcal{I}_1}{2}+\frac{R^2}{6}\,, \\
K_3 &:=
{}^*R_{{\alpha\beta \mu\nu}}^{*}R^{\alpha\beta \mu\nu}
= K-2\,\mathcal{I}_1-\frac{R^2}{3}\,.
\end{align}

Besides the standard polynomial curvature scalars, it is often useful to characterize the geometry through the full set of  Zakhary-McIntosh  invariants \cite{Carminati:1991ddy,Zakhary:1997xas,Santosuosso1998,Lan:2023cvz,Borissova:2026wmn}, which consists of seventeen elements constructed from the Ricci and Weyl tensors in  four-dimensional Lorentzian spaces. These quantities provide a systematic diagnostic of possible curvature singularities and of the algebraic structure  of  the gravitational field. Typically, the collection \eqref{eq:curvaturescalar}  constitutes a representative subset of these seventeen independent scalars \cite{Steane2021,Cherubini2002}, and is generally sufficient to capture spacetime regularity \cite{Simpson:2021dyo}.

However, as we demonstrate in  Appendix \ref{Appendix-0}, the  (orthonormal-frame) components of the Riemann tensor, as well as \emph{all}  curvature invariants, can be expressed as polynomials  with constant coefficients in the three functions $X(r),Y(r),$ and $Z(r)$ defined by
\begin{subequations}
\label{X-Y-Z-text}
  \begin{align}
    X(r) &:= \frac{B''(r)}{2}\,, \\
    Y(r) &:= \frac{B'(r)}{2r}\,, \\
    Z(r) &:= \frac{1-B(r)}{r^2}\,.
\end{align}  
\end{subequations}
Remarkably, we find that the Kretschmann scalar \eqref{eq:K_scalar} constitutes a positive-definite quadratic form in $X,Y,$ and $Z$, as it is given  by
\begin{align}\label{kretschmann-X-Y-Z}
K = 4(X^2 + 4 Y^2 + Z^2).
\end{align} 
Consequently, the  finiteness of  $X,Y,$ and $Z$ implies the finiteness of $K$, and hence  of all other algebraic curvature invariants, including the full Zakhary-McIntosh set.  This proves that  the Kretschmann scalar \emph{alone} is sufficient to fully determine the curvature regularity of     static, spherically symmetric metrics~\eqref{eq:metric_static}. Notably, this result also extends to  two-function settings with $ g_{rr} g_{tt}  \neq -1$, as we discuss  in Sec. \ref{Sec:2-function-setup}.

Given the above premises, let us  now focus on the conditions ensuring the absence of curvature singularities at an arbitrary ordinary point $r=r_0>0$ \footnote{By an ordinary point, we mean a point $r_0>0$ where the static coordinate chart is nondegenerate. For the one-function metric \eqref{eq:metric_static},  this amounts to
$0<|B(r_0)|<\infty$. Killing horizons $r=r_h$ are excluded from this class, since,  in Schwarzschild-like coordinates, $B(r_h)=0$, even though curvature invariants may remain finite there.}  (the asymptotic regime $r \to \infty$ will be considered in Sec. \ref{Sec:sufficient-conditions}). Introducing
\begin{equation}
  \Delta r:=r-r_0,
\end{equation}
we can write the local expansion
\begin{align}
B(r)
  &=
  b_0+b_1\Delta r+b_2\Delta r^2
  +\mathcal{O}(\Delta r^3),
  \label{eq:B-expansion-r0}
\end{align}
where $b_0\equiv B(r_0)$ is supposed to be finite
and nonzero. Plugging this formula into Eq.~\eqref{X-Y-Z-text} gives, to zeroth order in $\Delta r$, 
\begin{subequations} \label{eq:X-at-r0} 
\begin{align}
X(r)&=b_2+\mathcal{O}(\Delta r),
\\
Y(r)&=\frac{b_1}{2r_0}+\mathcal{O}(\Delta r),
\\
Z(r)&=\frac{1-b_0}{r_0^2}+\mathcal{O}(\Delta r).
\end{align}
\end{subequations}
As a consequence, at any ordinary point $r_0>0$,  the  Riemann tensor  components,  together with all algebraic curvature invariants, are bounded provided that metric function $B(r)$ is sufficiently differentiable and has finite, nonvanishing values at $r_0$. In particular, no negative powers of $\Delta r$ arise in Eq. \eqref{eq:X-at-r0}, and an expansion about $r_0>0$ imposes no additional algebraic restrictions on the series coefficients.

The center $r=0$ is geometrically distinguished from an ordinary point, because the symmetry two-spheres shrink to zero area there and the  functions
${Y}$ and ${Z}$ contain explicit inverse powers of the areal radius.  To determine the corresponding central regularity conditions, we assume that  $B(r)$ admits the power-law  behavior around $r = 0$
\begin{equation}\label{eq:B_expansion}
B(r) = B_0 + B_1 r + B_2 r^2 + B_3 r^3 + \mathcal{O}(r^4)\,,
\end{equation}
with $B_0 \equiv B(0)$. It then follows immediately  from identity \eqref{kretschmann-X-Y-Z} that
\begin{equation}\label{eq:B01}
B_0 = 1\,, \qquad B_1 = 0\,,
\end{equation}
are necessary and sufficient for the Kretschmann scalar to remain bounded at $r=0$. No restrictions on $B_2$ and $B_3$ arise at this stage, yet they will emerge once the energy conditions are taken into account, as we will see in Sec. \ref{Sec:role-ECS}.

\subsubsection{Static and spherically symmetric two-function geometries}\label{Sec:2-function-setup}

In this section, we  briefly discuss  the regularity properties of the  general two-function spacetime 
\begin{equation}
  \mathrm{d}s^2
  =
  -B(r)\,\mathrm{d}t^2
  +A(r)\,\mathrm{d}r^2
  +r^2\mathrm{d}\Omega^2,
  \label{eq:generic-static-metric}
\end{equation}
which  reduces to the metric \eqref{eq:metric_static} in the special case $A(r)B(r)=1$.

In this setting,  the independent nonvanishing components of the Riemann tensor can be encoded in the four quantities
\begin{align}\label{eq:X14-text}
    X_1(r):=\frac{(B'/\sqrt{AB})'}{2\sqrt{AB}} \,, \quad X_2(r):=\frac{B'}{2rAB} \,, \quad 
    X_3(r):=\frac{A'}{2rA^2} \,, \quad  X_4(r):=\frac{1-1/A}{r^2}\,,
\end{align}
and  the Kretschmann scalar takes the manifestly positive-definite form
\begin{equation}
K =  4(X_1^2+2X_2^2+2X_3^2+X_4^2). 
  \label{eq:generic-K-positive}
\end{equation}
Remarkably, all   curvature invariants constructed from the Riemann tensor and its contractions (and involving no covariant derivatives of the curvature) are polynomial functions  of the ${X}_i$ with constant coefficients. Therefore,   the boundedness of the Kretschmann scalar is sufficient to ensure the finiteness of all algebraic curvature invariants for the geometry \eqref{eq:generic-static-metric} as well. Further details are provided in Appendix~\ref{Appendix-K-ge}.

Similarly to Eq. \eqref{eq:B_expansion}, we can assess the regularity at $r=0$ by means of the expansions 
\begin{subequations}
\label{eq:generic-central-expansion}
\begin{align}
  A(r)
  &=
  A_0+A_1r+A_2r^2
  +\mathcal{O}(r^3),
  \label{eq:A-central-expansion}
  \\
  B(r)
  &=
  B_0+B_1r+B_2r^2
  +\mathcal{O}(r^3),
  \label{eq:B-central-expansion}
\end{align}
\end{subequations}
 which imply that near the center  
\begin{subequations}
\label{eq:generic-central-leading}
\begin{align}
  {X}_2(r)
  &=
  \frac{B_1}{2A_0B_0}\frac{1}{r}
  +\mathcal{O}(1),
  \\
  {X}_3(r)
  &=
  \frac{A_1}{2A_0^2}\frac{1}{r}
  +\mathcal{O}(1),
  \\
  {X}_4(r)
  &=
  \left(1-\frac{1}{A_0}\right)\frac{1}{r^2}
  +
  \frac{A_1}{A_0^2}\frac{1}{r}
  +\mathcal{O}(1),
\end{align}
\end{subequations}
while $X_1(r)$  contains only positive powers of $r$. 
Thus,  the necessary and sufficient conditions for the boundedness of all algebraic curvature scalars at $r=0$ are 
\begin{equation}
  A_0=1,
  \qquad
  A_1=0,
  \qquad
  B_1=0.
  \label{eq:generic-central-regularity}
\end{equation}
Interestingly,  unlike the Kerr-Schild scenario, no constraint  arises on  the value of $B_0$, and only its finiteness is required for the power series \eqref{eq:B-central-expansion} to be well defined.

\subsection{The role of energy conditions}\label{Sec:role-ECS}

Classical energy conditions configure as pointwise inequalities involving the components of the stress-energy tensor that embody physically motivated assumptions on matter fields entering Einstein field equations. They are  formulated as \cite{Curiel2017,Martin-Moruno2017}
\begin{align}
& T_{\mu \nu} k^{\mu} k^{\nu} \geq 0, \label{NEC-def-1}
 \\
& T_{\mu \nu} v^{\mu} v^{\nu} \geq 0 \,,
 \\
& \left(T_{\mu \nu} - \frac{1}{2}T\, g_{\mu \nu}\right) v^{\mu} v^{\nu} \geq 0 \,,
 \\
& T_{\mu\nu}v^\mu v^\nu \ge 0 \;\;\;  \text{and} \;\;\;  J^\mu J_\mu \le 0 \,, \label{DEC-def-1}
\end{align}
where $k^{\mu}$ is a null vector,   $v^\mu$ a normalized timelike vector,   $T=g^{\mu \nu}T_{\mu \nu}$ the trace of the stress-energy tensor, and   $J^\mu=-T^{\mu}{}_{\nu}v^\nu$ the energy flux vector. The first relation defines the  null energy condition (NEC) and represents the minimal requirement for the focusing of null geodesics, while the second, referred to as  weak energy condition (WEC), demands that the local energy density measured by any timelike observer be non-negative, thereby excluding the presence of exotic matter.  The third inequality corresponds to the strong energy condition (SEC), which  imposes that gravity is generically attractive, while the last one, the  dominant energy condition (DEC),  asserts that   energy propagation is causal. Accordingly, the following implications hold:
\begin{align}
& \text{WEC} \Rightarrow   \text{NEC},
\nonumber \\
& \text{SEC} \Rightarrow   \text{NEC},
\nonumber \\
& \text{DEC} \Rightarrow   \text{NEC}, \text{WEC}. 
\end{align}

The standard approach  to evaluating  energy criteria is to introduce the local orthonormal basis carried by a static observer. This frame can be constructed via the tetrad field $e^{a}{}_{\mu}$ \cite{MTW,Nakahara2003}, which constitutes a $4 \times 4$ matrix with positive determinant  satisfying  the completeness relation
\begin{align}
g_{\mu \nu} =e^{a}{}_{\mu}e^{b}{}_{\nu}\,\eta_{ab}\,,
\end{align}
with its inverse $e_a{}^{\mu}$ obeying the orthonormality condition
\begin{align}\label{eq:orthonormality condition}
g_{\mu \nu}\,e_{a}{}^{\mu}e_{b}{}^{\nu}= \eta_{ab} \,.
\end{align}
Here,   $\eta_{ab} = \mathrm{diag}(-1,1,1,1)$ is the Minkowski metric, and Latin  indices  label frame components.

When projected onto the static  frame, the energy-momentum tensor attains the diagonal form~\cite{Wald,Poisson2009}
\begin{align}\label{eq:diagT}
    T_{ab}= T_{\mu\nu} e_{a}{}^\mu e_{b}{}^\nu= {\rm diag} (\rho,p_1,p_2,p_3)\,,
\end{align}
where $\rho$ denotes the total  density of mass-energy, $p_1$  the radial pressure,   $p_2$ and $p_3$ the tangential pressures \cite{Morris-Thorne1988,Rezzolla-book2013}, and 
\begin{align}
e_{a}{}^{\mu} =
\begin{cases}
\mathrm{diag}\left[ 1/\sqrt{B},\, \sqrt{B},\, 1/r,\, 1/\left(r\sin\theta\right) \right], 
\quad  \;\;\; \;\;\,\, (B > 0), & \\
\mathrm{diag}\left[ 1/\sqrt{-B},\, \sqrt{-B},\, 1/r,\, 1/\left(r\sin\theta \right) \right],  \;\;\;\;\;  (B < 0), &
\end{cases}
\label{eq:orthonormal_frame}
\end{align}
which is a standard result for the  geometry  \eqref{eq:metric_static}. In this way, the classical energy relations \eqref{NEC-def-1}--\eqref{DEC-def-1} read as \cite{Poisson2009}
\begin{subequations}\label{eq:energycondition_rhoP}
   \begin{align}
    \text{NEC} &\Leftrightarrow \rho+p_i\geq0\,, \label{NEC-def}\\
     \text{WEC} &\Leftrightarrow \rho\geq 0, \;\rho+p_i\geq0\,, \label{WEC-def}\\
      \text{SEC} &\Leftrightarrow \rho+\sum_i p_i\geq0\,,\rho+p_i\geq0\,, \label{SEC-def}\\
      \text{DEC} &\Leftrightarrow \rho\geq 0, \;\rho\geq|p_i|\,, \label{DEC-def}
\end{align} 
\end{subequations}
for all $i=1,\,2,\,3$. We note that the diagonalization  of the energy-momentum tensor  may break down at a (Killing) horizon $r=r_h$ (where $B(r_h)=0$), since $e_{a}{}^{\mu}$ becomes singular. Nevertheless, the energy conditions at the horizon can be obtained as the limit  $r \to r_h$ of Eq.~\eqref{eq:energycondition_rhoP}, which is defined for $B \neq 0$ \cite{Maeda:2021ukk}. 

In the region where $B(r)>0$, the Einstein field equations yield
\begin{subequations}\label{eq:rhoPoutBH}
\begin{align}\label{eq:rho_B}
\rho &= \frac{1-(rB)'}{8\pi r^2}\,,\\
p_1 &= \frac{(rB)'-1}{8\pi r^2}\,,\\
p_2 &= p_3 = \frac{(rB)''}{16\pi r}\,,
\end{align}
\end{subequations}
with the same formulas holding also in the domain $B(r)<0$.
This coincidence relies  on the specific metric choice  $g_{rr}g_{tt}=-1$ underlying  Eq.~\eqref{eq:metric_static}.  By contrast,  for the generic  geometry~\eqref{eq:generic-static-metric}, the eigenvalue analysis of the energy-momentum tensor can still be carried out away from horizons and other degenerate surfaces, but  the identification of the ensuing functions  with the energy density and principal pressures depends on the causal character of the temporal and radial directions. Consequently, the explicit expressions for $\rho$ and $p_i$ generally differ between regions in which $B(r)>0$ and $B(r)<0$. 
In particular, in a connected static domain with $A(r)>0$ and $B(r)>0$, the radial NEC implies~\cite{Wang-Battista2026a}
\begin{equation} 
8\pi\left(\rho+p_1\right) = \frac{\left[ A(r)B(r) \right]'} {A(r)^2B(r)r} \geq 0,
\label{eq:generic-radial-NEC} 
\end{equation} 
so that  $A(r)B(r)$ must be nondecreasing. The one-function subclass $A(r)B(r)=1$ identically saturates the above inequality, yielding $\rho+p_1=0$.
A detailed examination of the diagonalization of the energy-momentum tensor and of the resulting energy conditions for generic static, spherically symmetric spacetimes was presented in our previous work~\cite{Wang-Battista2026a}. That study also addresses the subtleties arising at Killing horizons and in configurations  where $A(r)B(r)$ can vanish.

Returning to the Kerr-Schild framework, in the special isotropic-pressure scenario, corresponding to  $p_1=p_2$  in Eq.~\eqref{eq:rhoPoutBH}, the  function $B(r)$ reduces to
\begin{align}\label{eq:metric_iso}
    B(r)=1-\frac{c_1}{r}+{c_2}r^2\,,
\end{align}
and the  energy criteria \eqref{eq:energycondition_rhoP} imply   
\begin{align}
\rho=-\frac{3c_2}{8\pi}\,,\qquad
p_1=p_2=p_3=-\rho\,, 
\end{align}
with $c_{1}$ and $c_{2}$  arbitrary constants.  Consequently,  the NEC  holds identically, while the WEC and DEC  call simply for  $c_2\le 0$. Enforcing  regularity at $r=0$ further sets $c_1=0$, which results in  a  solution of either  de Sitter type (for $c_2<0$) or Minkowski type (for $c_2=0$).

In the general  configuration  with anisotropic pressure,  starting from the near-origin series \eqref{eq:B_expansion},  we find that the energy conditions lead to the constraints 
\begin{align}
   B_0 = 1,\qquad B_1 = 0,\qquad B_2 = 0,\qquad B_3 = 0\,. 
\end{align}
In this case,  one can show that the WEC and  SEC cannot be simultaneously fulfilled for any nontrivial power-law expansion of $B(r)$ near $r=0$.  The argument proceeds as follows. Suppose that the leading nonvanishing contribution to $B(r)$ takes the form
\begin{equation}
B(r)=1+B_n r^n\,, 
\end{equation}
where we set $n\ge 2$ in view of the regularity relations \eqref{eq:B01}. Bearing in mind Eq. \eqref{eq:rhoPoutBH},  the WEC entails $\rho\ge 0$,  immediately yielding
\begin{equation}\label{eq:wecrq}
B_n < 0\,,
\end{equation}
while the SEC demands  the tangential pressure   $p_2$ to be positive in a neighborhood of the origin, which amounts to
\begin{equation}
B_n > 0\,,
\end{equation}
in  clear contradiction with Eq.~\eqref{eq:wecrq}.
Therefore, the WEC and SEC cannot be jointly satisfied, as anticipated before. This result is fully consistent with Hawking-Penrose theorems~\cite{Penrose:1964wq,Hawking:1970zqf}, which indicate that  avoiding  singularities in regular spacetimes necessarily forces the violation of at least one of the classical energy conditions~\cite{Curiel2017} \footnote{There exist alternative scenarios, such as defect spacetimes~\cite{Klinkhamer:2013wla,Klinkhamer:2018xot}, where curvature singularities may be avoided without violating the standard energy conditions. These constructions usually involve degenerate metrics and thus lie beyond the scope of classical general relativity (see also Refs. \cite{Klinkhamer2019a,Battista2020defect} for cosmological applications). We do not consider such cases in this work.}. For this reason, we hereafter restrict our attention to  the NEC and WEC  and search for families of regular spacetimes compatible with them.

\subsection{Sufficient conditions for regular spacetimes under the WEC}\label{Sec:sufficient-conditions}

To construct a regular spacetime satisfying the NEC and WEC, the following requirements must be fulfilled for any  $r \geq 0$:
\begin{subequations}\label{eq:requirementwec}
\begin{align}
   & \rho(r) \geq 0 \ \text{and finite}, \label{eq:requirementwec-1}\\[2pt]
   & \rho(r) + p_i(r) \geq 0 \ \text{and finite}, \label{eq:requirementwec-2} \\[2pt]
   & K(r) \;\text{remains finite}, \label{eq:requirementwec-3}
\end{align}
\end{subequations}
the latter constraint reflecting the regularity analysis carried out in Sec. \ref{Sec:curvature-invariants} (and Appendix \ref{Appendix-0}).

To broaden the scope of the analysis,  we no longer assume the near-zone expansion \eqref{eq:B_expansion}. Instead,  Eq. \eqref{eq:rho_B} suggests   introducing the mass function~\cite{Wald}
\begin{equation}
m(r) \equiv 4\pi \int_0^r \rho(\tilde r)\,\tilde r^2\, \dd \tilde r ,
\label{eq:mass_function}
\end{equation}
which allows  the metric component $B(r)$ to be written, in full generality, as 
\begin{equation}
B(r) = 1 - \frac{2m(r)}{r}.
\label{eq:B_mass}
\end{equation}
Here, $m(r)$ equals the Misner-Sharp-Hernandez mass,  a key quantity for assessing the  asymptotic flatness of spherically symmetric configurations  defined as $\mathscr{M}:=\left(\tfrac{r}{2}\right)\left(1-g^{\mu \nu} \nabla_\mu r \nabla_\nu r\right)$ \cite{Faraoni2021}. Therefore, using identity \eqref{kretschmann-X-Y-Z} jointly with Einstein-equation relations  \eqref{eq:rhoPoutBH}, we find  
\begin{align}
&\rho(r) + p_1(r)=0, \label{NEC-WEC-rel1}\\
&\rho(r) + p_2(r) = -\frac{r}{2}\rho'(r)=\rho(r) + p_3(r), \label{NEC-WEC-rel2}\\
& K(r) =
\frac{48\,m(r)^2}{r^6}
- \frac{64\pi\,m(r)}{r^3}\bigl[2\rho(r) - r\rho'(r)\bigr]
+ 64\pi^2\bigl[4\rho(r)^2 + r^2\rho'(r)^2\bigr]\,. \label{NEC-WEC-rel4} 
\end{align}
It then follows from relations \eqref{eq:requirementwec} that a set of \emph{sufficient} conditions to construct a regular Kerr-Schild spacetime satisfying the WEC (and hence the NEC)  is
\begin{subequations}
\label{eq:requirerho}
\begin{align}
\rho(r)& \geq 0, \label{eq:requirerho-2}\\
\rho'(r) & \le 0,   \label{eq:requirerho-3}\\
\rho(r) &\in C^1. \label{eq:requirerho-1}
\end{align}
\end{subequations}
Regarding the remaining energy criteria, the DEC is not enforced at this stage, since it would introduce the additional inequality $\rho + r\rho'/4 \geq 0$, which   cannot be straightforwardly incorporated within our framework; the SEC is likewise excluded, as we have already shown  that it cannot hold simultaneously with the WEC. The energy density is thus the central quantity of our framework, since once it is specified, the pressures are completely determined by Eqs.~\eqref{NEC-WEC-rel1} and \eqref{NEC-WEC-rel2}. In particular,  $p_2=p_3=-\rho-r\rho'/2 $, so  that the  contribution of the tangential pressures to the energy criteria is entirely controlled by the derivative of the density profile.   

Assumptions \eqref{eq:requirerho-2} and \eqref{eq:requirerho-3}  are necessary and sufficient for the validity of the WEC, while Eq. \eqref{eq:requirerho-1} is sufficient for the spacetime  regularity  at any finite radius, and entails  that the functions \eqref{X-Y-Z-text}, and consequently all curvature invariants, remain bounded for any  finite $r$. In particular, since   $\rho(r)$ is supposed to be continuously differentiable, it  admits the first-order expansion  $\rho(r) = \rho(0) + \rho'(0)\,r + o(r)$  as $r\to 0^+$. Substituting this into the definition \eqref{eq:mass_function} of $m(r)$, yields 
\begin{align}\label{eq:m0}
m(r)
=4\pi\int_0^r \left[\rho(0)+  \rho'(0)\tilde r+ o(\tilde r)\right]\tilde r^2\,\mathrm{d}\tilde r
=\frac{4\pi}{3}\rho(0)\,r^3+o(r^3).
\end{align}
Hence, by the definition of little-$o$, 
\begin{align}
\lim_{r\to0}\frac{m(r)}{r^3}
=\frac{4\pi}{3}\rho(0),    \label{limit-little-o}
\end{align}
which is finite under our hypotheses. The quantity $m(r)/r^3$ is thus regular at the center.

To restrict attention to physically reasonable scenarios, we further take into account the two
additional assumptions
\begin{subequations}\label{eq:addition_all}
\begin{align}
\rho'(0)&=0\,,   \label{eq:addtion_req-2} \\
\lim_{r\to\infty} B(r)&=1 \,. \label{eq:addtion_req}
\end{align} 
\end{subequations}
The former  accounts for a flat density profile at the core. This is a necessary condition for a static, locally isotropic matter distribution  to maintain mechanical equilibrium at the origin and, in turn,  naturally leads to a maximally symmetric geometry (locally de Sitter or Minkowski) \cite{Fan:2016hvf}, described near $r=0$ by Eq. \eqref{eq:metric_iso} with $c_1=0$.
Relation \eqref{eq:addtion_req} ensures that the metric  approaches the Minkowski form asymptotically  so that curvature scalars remain finite at spatial infinity. This means that, under our hypotheses,  curvature invariants do not diverge  anywhere in the spacetime. 

As first shown in Ref.~\cite{Hu:2023iuw},  regularity may be characterized through the absence of curvature singularities, as this notion is equivalent to geodesic completeness for   Kerr-Schild geometries~\eqref{eq:metric_static} possessing a regular core and strict asymptotic flatness (namely, configurations for which the Misner-Sharp-Hernandez mass $m(r)$ approaches a finite nonvanishing value at infinity). We can now see that this result can be generalized, as Eq.~\eqref{eq:addtion_req} is sufficient for the divergence argument underlying the completeness of null geodesics. Indeed, using Gullstrand-Painlev\'e coordinates $(t_\ast,r,\theta,\phi)$, one  easily finds~\cite{Hu:2023iuw}   that the time $t_\ast$ associated with an ingoing radial light ray traveling from spatial infinity down to the center $r=0$ is 
\begin{align}
    t_\ast=-\int_\infty^0 \frac{\dd r}{1+\sqrt{2m(r)/r}}.
\end{align}
The quantity $t_\ast$ blows up to $-\infty$ whenever $m(r)/r \to 0 $ as $r \to \infty$, or equivalently when  Eq.~\eqref{eq:addtion_req} holds. Indeed, in this limit  the term $\sqrt{2 m(r) /r}$  becomes negligible, so that the integrand approaches unity at large $r$ and the integral diverges linearly. In addition, completeness of timelike geodesics  stems  from core regularity~\eqref{limit-little-o}, as proved in Ref.~\cite{Hu:2023iuw}. Therefore, the equivalence between causal geodesic completeness and the finiteness of curvature scalars extends beyond the strictly asymptotically flat scenario and remains valid   also for metrics approaching the Minkowski form at infinity. This applies, in particular, to all regular solutions considered in this paper.

We can thus conclude that the system of ordinary differential equations~\eqref{NEC-WEC-rel1}--\eqref{NEC-WEC-rel4} provides a systematic algorithm for deriving families of regular solutions adhering to the WEC: once   a density profile  $\rho(r)$ obeying Eq.~\eqref{eq:requirerho} is given, the corresponding metric component $B(r)$ follows directly from solving  Eq.~\eqref{eq:rho_B}. Explicit applications of this scheme will be the main focus of the next section.

\section{Families of regular spacetimes from admissible density profiles}
\label{sec:3}

In this section, we employ the algorithm introduced before and explore various regular spacetimes adhering to  WEC and   sourced by a mass-energy density $\rho(r)$  complying with relations \eqref{eq:requirerho} and \eqref{eq:addtion_req-2}. The different cases  analyzed admit the general parametrization
\begin{equation}
\rho(r)=\rho_0\,F(r)\,,
\label{density-parametrization}
\end{equation}
where $\rho_0\equiv \rho(0)$ denotes the central density and $F(r)$ is a dimensionless shape function  containing a characteristic length scale  $r_d$,  normalized so that $F(0)=1$. We therefore set aside  regular scenarios with $F(0)=0$ (which have also been proposed in the literature   \cite{Simpson:2019mud,Simpson:2021dyo}), since  they lead to   Minkowski-core configurations where typically $\rho'(r)>0$ over some interval of $r$, thereby violating the NEC (cf. Eq. \eqref{eq:requirerho-3}). 

To organize the admissible profiles in a systematic way,  we introduce a parameter $\mathcal{C}$ that quantifies  the functional complexity of $\rho(r)$. This quantity is not intended to establish a unique or universal definition of complexity, but rather  provides a practical  tool for classifying the  parametric forms of $\rho(r)$  according to their structural simplicity. In this way, we can study  regular geometries  starting from minimally complex densities  and gradually proceeding to more elaborate cases.

We define the factor $\mathcal{C}$ as 
\begin{equation}\label{eq:complexityC}
\mathcal{C} := N_{\rm param} + N_{\rm op} + \alpha \, N_{\rm comp},
\end{equation}
where
\begin{itemize}
\item $N_{\rm param}$ denotes the number of free,  dimensionless \emph{shape parameters} that affect the functional form of $\rho(r)$, excluding the overall normalization  $\rho_0$ and the length scale $r_d$; 
\item $N_{\rm op}$ counts the number of  elementary operations required to  construct $\rho(r)$, including addition, subtraction, multiplication, division, and algebraic power operations;
\item $N_{\rm comp}$ tallies the number of \emph{nontrivial} elementary function compositions, such as $\exp(f)$ or $\ln(1+f)$, where $f=f(r)$;
\item $\alpha$ is a weighting factor controlling the relative importance of compositions, and we set $\alpha=2$ for simplicity.
\end{itemize}

\begin{table}[h]
\centering
\begin{tabular}{l l c c c c}
\toprule
Category 
& Function $F(x)$  
& $N_{\rm param}$ 
& $N_{\rm op}$ 
& $N_{\rm comp}$ 
& $\mathcal{C}$ \\
\midrule
\multirow{4}{*}{\shortstack{Examples of \\basic functions}}
& ${1}/{x}$ & 0 & 0 & 0 & 0 \\
& ${1}/{x^2}$ & 0 & 1 & 0 & 1 \\
& $\left(1 + x^2\right)^{-1}$ & 0 & 2 & 0 & 2 \\
& $\exp(-x)$ & 0 & 0 & 1 & 2 \\
\midrule
\multirow{5}{*}{\shortstack{Shape functions \\ studied in this paper}}
& $\left(1 + x^n\right)^{-1}$ & 1  & 2 & 0 & 3 \\
& $\exp(-x^n)$& 1  & 1 & 1 & 4 \\
& $\left(1 + x^n\right)^{-\ell}$ & 2  & 3 & 0 & 5 \\
& $x^{\ell} \exp(-x^n)$ & 2  & 3 & 1 & 7 \\
& $x^{\kappa}\left(1 + x^n\right)^{-\ell}$& 3  & 5 & 0 & 8 \\
%& $(1+x^{m})(1 + x^n)^{-\ell}$ & 3 ($m,n,\ell$) & 6 & 0 & 9 \\
\bottomrule
\end{tabular}
\caption{Shape functions $F(x)$ (with $x:=r/r_{d}$) and their associated functional complexity $\mathcal{C}$.  Since $\mathcal{C}$ is designed to capture the structural hierarchy of functional forms, algebraically equivalent expressions, including sign changes and reciprocal representations, are assigned the same  complexity factor.  Accordingly, we set $N_{\rm op}=0$ for $F(x)=1/x$, so that  $F(x)=x^{n}$ and $F(x)=x^{-n}$ carry the same value of  $N_{\rm op}$, and the operation count does not depend on the sign of the shape parameter $n$. }
\label{tab:complexity_examples}
\end{table}

Representative examples of functions, along with their corresponding complexity parameter, are listed in Table~\ref{tab:complexity_examples}, which  illustrates how  increasingly elaborate functional structures naturally correspond to larger values of $\mathcal{C}$. As an example, the function $F(x)=\left(1 + x^n\right)^{-1}$ (with $x:=r/r_{d}$) has  $\mathcal{C}=3$: it contains  the dimensionless exponent $n$ ($N_{\rm param}=1$),  involves two elementary operations (one addition and one power, so  $N_{\rm op}=2$), and includes no trivial compositions ($N_{\rm comp}=0$). Likewise, $F(x)=x^{\ell} \exp(-x^n)$ has $\mathcal{C}=7$: two shape parameters $\ell$ and $n$ ($N_{\rm param}=2$), three elementary operations (two power operations and one multiplication and hence $N_{\rm op}=3$), and one composition  (the exponential function, which yields $N_{\rm comp}=1$). 
Of course, many other possibilities exist beyond  those listed  in Table~\ref{tab:complexity_examples}. For instance, regular black holes studied in Ref.~\cite{Ayon-Beato:1999kuh} adopt hyperbolic density functions that spoil  the NEC at the origin, and will therefore not be discussed here.  However, in such a setup, the typical functions encountered are of the form $F(r)\sim r^{-n} {\rm sech}^2(r_d/r) $, for which one typically finds $\mathcal{C}>10$. These profiles are therefore to be regarded as highly complex, when expressed in terms of elementary functions.

Guided by the admissibility  criteria \eqref{eq:requirerho} and \eqref{eq:addtion_req-2}, we now investigate energy-density profiles in order of  increasing values of $\mathcal{C}$, together with  the ensuing geometries. We begin with the simplest rational-falloff forms (Sec. \ref{Sec:rational-falloff}), then  consider  power-law functions  (Sec. \ref{sec:rhopower5}), exponentially suppressed densities  (Sec. \ref{sec:exp1}), and finally composite constructions  combining algebraic and exponential structures (Sec. \ref{sec:2.5}). We conclude the section by discussing the possibility of superposing the resulting solutions (Sec. \ref{sec:5}). For all  examined configurations,  compatibility with the WEC (or even the DEC) can be easily determined by inspecting the allowed ranges of the free parameters of the model. As we will see, although  several of the density functions  considered below have already appeared in the literature, particularly in the  studies of dark matter (strictly speaking, the only apparent exception is  the simplest scenario with shape function $F(x)=\left(1 + x^n\right)^{-1}$, which does not seem to have
been singled out and systematically analyzed), their implementation within the proposed WEC-based construction generates novel families of regular spacetimes.

\subsection{Rational-falloff density profiles: $\rho(r) = \rho_0  \left[1+(r/r_d)^n\right]^{-1}$}\label{Sec:rational-falloff}

According to our definition \eqref{eq:complexityC} of functional complexity, the simplest  form assumed by  $\rho(r)$ corresponds to the rational power-law falloff
\begin{align}\label{eq:rho_asantz0}
    \rho(r) = \frac{\rho_0}{1+(r/r_d)^n}\,.
\end{align}

The relations \eqref{eq:requirerho}, and consequently the WEC, are obeyed whenever $n \geq 1$. First, given the non-negativity assumption \eqref{eq:requirerho-2}, we take the $r=0$ density $\rho_0$ to be  positive (and  finite). For $n>0$, the monotonicity   constraint \eqref{eq:requirerho-3} is then automatically fulfilled, while the smoothness criterion \eqref{eq:requirerho-1} calls for $n \geq 1$.  This bound becomes more stringent once the  central isotropy condition \eqref{eq:addtion_req-2} is imposed, which entails  $n>1$. Enforcing the DEC  the additional inequality $n\leq 4$ arises, which will be relaxed in the following  to keep the discussion as general as possible. 

The asymptotic flatness of the spacetime can be   examined through  the large-$r$ behavior of $\rho(r)$, which  scales as $\rho(r) \sim \rho_0(r_d/r)^n$ for $r\gg r_d$. Substituting this  into the   formula \eqref{eq:mass_function}  for the   total enclosed mass yields 
\begin{equation}\label{eq:mass1}
m(r) \sim  \rho_0 \, (r_d)^{n}\int r^{2-n} \, \mathrm{d}r , 
\end{equation}
and, accordingly, four distinct regimes can be identified, depending on the value of  $n$:
\begin{itemize}
\item \textbf{$n>3$:}
$m(r)$ converges as $r\to\infty$. This situation points to a finite ADM mass and a strictly asymptotically flat geometry.  
\item \textbf{$n=3$:} 
the mass function grows  logarithmically, and hence the ADM mass fails to converge to a finite value at infinity. We thus find   what we may call a \qm{\emph{marginally} asymptotically flat spacetime}, meaning that  the metric approaches the Minkowski form at large $r$ (i.e., it complies with Eq. \eqref{eq:addtion_req}), but with a falloff slower than $1/r$. As we will see in Sec. \ref{Sec:n=3,6-hyper}, such  behavior generically leads to  geometries involving logarithmic corrections  (see also Refs. \cite{Wang-Battista2026a,Liang:2023jrj}). 
\item \textbf{$2<n<3$:} 
$m(r)$ diverges as a power law, but more slowly than the linear function $r$. This  again corresponds to a marginally asymptotically flat scenario.  
\item \textbf{$n\leq2$:} 
$m(r)$ blows up with a growth rate equal to or faster than $r$, indicating a non-asymptotically flat solution. Such cases will not be considered in the present work. 
\end{itemize}

With the \emph{ansatz}~\eqref{eq:rho_asantz0}, the metric function $B(r)$, obtained by solving Eq.~\eqref{eq:rho_B}, can be formulated in terms of the Gauss hypergeometric function 
$F(a,b;c;z)\equiv {}_2F_1(a,b;c;z)$~\cite{abramowitz1968handbook}.
Explicitly, we find
\begin{align}\label{eq:B_asantz0}
  B(r)= 1+\frac{c_1}{r}
  - \frac{ r^2 \lambda}{3r_d^2}\,
 F\!\left[1,\frac{3}{n};1+\frac{3}{n};-\left(\frac{r}{r_d}\right)^n\right],
\end{align}
with $c_1$ an integration constant and  $\lambda := 8\pi \rho_0 r_d^2$\,  a  dimensionless quantity. One can show that the Minkowski limit \eqref{eq:addtion_req} is achieved for $n>2$,  in agreement with our previous  analysis. In this regime, spacetime regularity can  equivalently be accommodated either by the boundedness of the Kretschmann scalar \eqref{kretschmann-X-Y-Z} or by the completeness of causal geodesics. 

From the above result, we also see that the only freedom in the algorithm devised in Sec. \ref{Sec:sufficient-conditions} for   constructing   regular WEC-respecting solutions concerns the  constant $c_1$. Whether $c_1$
vanishes or not depends on the behavior of $B(r)$ near the origin (for instance, we will see that it is zero for the logarithmic solution~\eqref{eq:sol_ln0}, while remaining nonzero for the black hole geometry~\eqref{eq:Bsol_exp_n3}).

For specific values of the parameter $n$, the hypergeometric function admits a representation in terms of elementary  contributions. The special cases  $n=3$ and $n=6$ are analyzed in Sec.~\ref{Sec:n=3,6-hyper}; to illustrate a representative generic scenario, the configuration with  $n=4$ is studied  in Sec. \ref{Sec:n-neq-3-6-hyper}.

\subsubsection{Logarithmic and arctan-modified geometries ($n=3$ and $n=6$)}\label{Sec:n=3,6-hyper}

In the situation with $n=3$, the hypergeometric function entering Eq.~\eqref{eq:B_asantz0} takes the closed-form expression \cite{abramowitz1968handbook}
\begin{align}
F(1,1;2;z)= -\frac{\ln(1-z)}{z},
\end{align}
which allows us to write 
\begin{align}\label{eq:B_asantz0_1}
  B(r)= 1+\frac{c_1}{r}
  -\frac{\lambda r_d}{3r}\ln\left(\frac{r^3}{r_d^3}+1\right).
\end{align}
The small-$r$ expansion 
\begin{align}
B(r)
= 1 + \frac{c_1}{r}
  - \frac{\lambda}{3r_d^2} r^2
  + \mathcal{O}\!\left(r^{5}\right), 
\end{align}
shows that regularity at the origin enforces $c_1=0$, thereby leading to a  spacetime with a central de Sitter core.  Therefore,  we end up with the  regular and  marginally asymptotically flat solution  
\begin{align}\label{eq:sol_ln0}
    B(r)=1-\frac{\lambda r_d}{3r}\ln\left(\frac{r^3}{r_d^3}+1\right)\,.
\end{align}

For $n=6$, we can exploit  the identity \cite{abramowitz1968handbook}
\begin{equation}
F\!\left(1,1/2;3/2;-z^2\right)=\frac{\arctan z}{z}\,, 
\end{equation}
to recast Eq. \eqref{eq:B_asantz0} as
\begin{equation}\label{eq:sol_arctan}
B(r)=1-\frac{\lambda r_d}{3r}\arctan \left(\frac{r^3}{r_d^3}\right)\,,
\end{equation}
where like before we have set $c_1=0$. In the asymptotic region $r\gg r_d$, we thus find 
\begin{equation}
B(r)\sim 1-\frac{2m_{\rm eff}}{r}\,, \label{m-eff-relation}
\end{equation}
where  $m_{\rm eff}$ denotes the effective ADM  mass, here   given by
\begin{equation}\label{m-eff}
m_{\rm eff}=\frac{\pi \lambda r_d}{12}\,.
\end{equation}

As far as we are aware, Eqs.~\eqref{eq:sol_ln0} and \eqref{eq:sol_arctan} have not been reported previously in the literature. The former represents the lowest-complexity  logarithmic  regular solution, while the latter is the simplest realization of an $\arctan$-based strictly asymptotically flat  geometry. Similar $\arctan$-modified models arising from a phantom scalar field were  presented in Ref.~\cite{Bronnikov:2005gm}, although in a more involved metric representation.

\subsubsection{Mixed arctan and logarithmic geometries  ($n=4$)} \label{Sec:n-neq-3-6-hyper}

Leaving aside the special cases $n=3$ and $n=6$, for generic $n \geq 2$ the hypergeometric function occurring in Eq. \eqref{eq:B_asantz0} does not admit a compact closed form in terms of a single elementary function, but can still be reduced to a finite combination of elementary contributions, typically involving logarithmic and inverse-trigonometric terms. In this situation, the basic strategy for evaluating $F\left[1,3/n;1+3/n;-\left(r/r_d\right)^n\right]$ is to start from its integral representation \cite{abramowitz1968handbook} and then exploit a partial-fraction decomposition. 

As an   example, let us consider  the parameter choice $n=4$, where one obtains
\begin{subequations}
\label{eq:Fn4}
\begin{align}
F\left[1,3/4;7/4;-(r/r_d)^4\right]
= \frac{3 r_d ^3}{r^3}I(r) \,,
\end{align}
\text{with}
\begin{align}
   I(r):=& \int_0 ^{r/r_d} \frac{u^{2}}{1+u^{4}} \dd u=   \frac{1}{2\sqrt{2}}\Bigl\{
\arctan\!\bigl[\sqrt{2}\,(r/r_d)+1\bigr]
+\arctan\!\bigl[\sqrt{2}\,(r/r_d)-1\bigr]
\Bigr\} \notag \\
&
+\frac{1}{4\sqrt{2}}\,
\ln\!\left[\frac{(r/r_d)^{2}-\sqrt{2}\,(r/r_d)+1}{(r/r_d)^{2}+\sqrt{2}\,(r/r_d)+1}\right]\,.\label{eq:defI}
\end{align}
\end{subequations}

Therefore, upon exploiting Eq.~\eqref{eq:B_asantz0} with $c_1=0$, we obtain  the novel  regular solution 
\begin{align} \label{B-and-I}
B(r) = 1 -\frac{\lambda r_d}{r} I(r)\,, 
\end{align}
which gives rise to a strictly asymptotically flat spacetime  featuring an effective ADM mass $m_{\rm eff} = \pi \lambda r_d  /(4\sqrt{2})$ and  a de Sitter core.

For other values of $n$, the resulting geometries can also be regular, but they generally come with  more complicated combinations of logarithmic and inverse-trigonometric functions. We shall not pursue these configurations further in this paper.

\subsection{Power-law density profiles: $\rho(r) = \rho_0 \left[1+(r/r_d)^n\right]^{-\ell}$} \label{sec:rhopower5}

In this section, we deal with a mass-energy density with power-law falloff
\begin{equation}\label{eq:rho_asantz1}
\rho(r) = \frac{\rho_0}{\left[1+(r/r_d)^n\right]^{\ell}},
\end{equation}
which represents an increase in functional complexity relative to the \emph{ansatz} \eqref{eq:rho_asantz0} (in this case $\mathcal{C}=5$, see Table \ref{tab:complexity_examples}). 
Owing to Eq. \eqref{eq:requirerho-2}, we take  $\rho_0> 0$  (and  finite), while the requirement of monotonic decrease  \eqref{eq:requirerho-3} furnishes  $n \ell>0$ (excluding the trivial setup of a constant density).
The profile is continuous  at $r=0$ for any positive $n$, and belongs to the class $C^1$, as mandated by Eq. \eqref{eq:requirerho-1},  provided   $n \geq 1$.  Thus, consistently with the  sufficient criteria  \eqref{eq:requirerho},  the WEC is satisfied whenever $n \geq 1$ and $\ell>0$.  If, in addition, the flat-core condition \eqref{eq:addtion_req-2} is imposed,   the parameter space is further limited  to
\begin{subequations}
\begin{align}
&n>1, \label{n-l-inequality-1}
\\
& \ell>0 . \label{n-l-inequality-2}
\end{align}
\end{subequations}
The DEC entails  the tighter bound  $n\ell \le 4$, which however we do not enforce to retain generality in our analysis.

The asymptotic behavior of the Misner-Sharp-Hernandez mass  (cf. Eq. \eqref{eq:mass_function}) is
\begin{equation} \label{eq:mass5}
m(r) \sim \rho_0 \left(r_d\right)^{n\ell}\int  r^{2-n\ell}\, \mathrm{d}r, 
\end{equation}
and hence the corresponding classification is  completely analogous to that obtained in Sec.~\ref{Sec:rational-falloff} (see the text below Eq.~\eqref{eq:mass1}), with the replacement $n\rightarrow n\ell$. In particular, the spacetime is strictly asymptotically flat for $n\ell>3$, marginally asymptotically flat for $2<n\ell\leq 3$, and non-asymptotically flat for $n\ell\leq 2$.

It is not complicated to see that the density profile \eqref{eq:rho_asantz1} leads to the metric 
\begin{equation}\label{eq:B_asantz1}
B(r)=1+\frac{c_1}{r}-\frac{\lambda r^2}{3r_d^2}\,
F\!\left[\ell,\frac{3}{n};1+\frac{3}{n};-\left(\frac{r}{r_d}\right)^n\right],
\end{equation}
which exhibits  the Minkowski asymptotics  \eqref{eq:addtion_req} for   $n\ell>2$, consistently with the above discussion.  

We notice that a further generalization of the power-law configuration  \eqref{eq:rho_asantz1} is
\begin{align}\label{generalized-profile-2}
\rho(r)=\rho_0 \left(\frac{r}{r_d}\right)^{\kappa}
\left[1 + \left(\frac{r}{r_d}\right)^n\right]^{-\ell},
\end{align} 
which features the additional term  $\left(r/r_d\right)^{\kappa}$ and represents the most complex scenario considered in this work, with $\mathcal{C}=8$ (see  Table \ref{tab:complexity_examples}).  Such profiles play an important role  within the context of dark-matter halo modeling \cite{Dehnen:1993uh,Navarro:1995iw,Hernquist:1990be,Zhao:1995cp} and  have also been exploited  recently  in the construction of regular black holes~\cite{Kar:2025phe,Konoplya:2025ect}. It is easy to see that the power-law prefactor must be absent, namely $\kappa=0$, in order to obtain a regular solution satisfying the WEC and possessing a finite, nonvanishing central density. With this choice, the model reduces to the class \eqref{eq:rho_asantz1}. 

For this reason, below we focus on the profile \eqref{eq:rho_asantz1} and consider particular choices of the parameters $n$ and $\ell$ for which the hypergeometric function boils down to elementary contributions. Explicitly, we  explore the two cases $\ell = 1+3/n$,   $n=3$ in Sec. \ref{Sec:ell-n-choice-1}, and $n=3/2$ in Sec. \ref{Sec:ell-n-choice-3}. Finally, in Sec.  \ref{Sec:ell-n-choice-4}, we examine solutions with even values of $n$ and present the analytic structure of the corresponding geometries for $n=2,4,6$.

\subsubsection{Bardeen,  Hayward, and other regular black holes ($\ell = 1+3/n$ or  $n=3$)}\label{Sec:ell-n-choice-1}

The first interesting setup arises for $\ell = 1 + 3/n$. By virtue of the identity $F(a,b;a;z) = (1-z)^{-b}$~\cite{abramowitz1968handbook},  we find that Eq. \eqref{eq:B_asantz1} boils down to
\begin{equation}\label{eq:B_sol1}
B(r)
= 1 
- \frac{\lambda r^2}{3r_d^2}\,
\left[1+\left(\frac{r}{r_d}\right)^n\right]^{-3/n}, 
\end{equation}
where regularity at the origin fixes $c_1=0$, yielding a  de Sitter core. The resulting configuration embodies a class of strictly asymptotically flat geometries with effective ADM mass  
\begin{align}
m_{\rm eff} = \frac{\lambda r_d}{6},
\end{align}
provided the lower bound $n>1$ is taken into account, which guarantees that $n\ell >3$. 

The geometry \eqref{eq:B_sol1} allows us to recover a  class of regular black hole solutions  well known in the literature~\cite{Fan:2016hvf,Kar:2025phe,Neves:2014aba}. In particular, our expression coincides with Eq.~(3.20) of Ref.~\cite{Kar:2025phe} upon setting $\mu=3$ therein \footnote{This correspondence does not imply that our construction is less general than that of Ref.~\cite{Kar:2025phe}. Here, we restrict attention to regular spacetimes satisfying the WEC, whereas Eq.~(3.20) of Ref.~\cite{Kar:2025phe} does not in general fulfill this requirement.}.  Moreover, the cases $n=2$ and $n=3$ reproduce the functional forms of the Bardeen~\cite{bardeen1968non} and Hayward~\cite{Hayward2005} regular black holes, respectively, after appropriate identifications of the parameters.

Let us now consider the scenario with $n=3$ and $\ell>1$, which, in view of our analysis, produces a strictly asymptotically flat configuration. In this situation, by using the general integral representation of the hypergeometric function~\cite{abramowitz1968handbook}, we  obtain from Eq.~\eqref{eq:B_asantz1} 
\begin{equation}
B(r) = 1-\frac{\lambda r_d}{3(\ell-1)r}
+ \frac{\lambda r_d}{3(\ell-1)r}\,
\left(1+\frac{r^3}{r_d^3}\right)^{-\ell+1},
\label{reg-geom-ell}
\end{equation}
where, again, we have set $c_1=0$ to ensure a regular  center. Furthermore,  the  large-$r$ expansion of $B(r)$ results in the effective ADM mass
\begin{align}
    m_{\rm eff} = \frac{\lambda r_d}{6(\ell-1)}.
\end{align}

The family of regular spacetimes~\eqref{reg-geom-ell} has also appeared in the literature. For instance, as shown in  Ref.~\cite{Konoplya:2025ect}, this metric  can describe  a black hole sourced by a specific   dark-matter halo model for suitable choices of the free parameters (see Eq.~(28) therein). In addition, consistently with the previous solution, the case $\ell=2$ again coincides with the Hayward black hole~\cite{Hayward2005}. 

It should be noted that one may also take $2/3<\ell<1$ in Eq.~\eqref{reg-geom-ell}. However, in this range, the spacetime is only marginally asymptotically flat. 

\subsubsection{The case $n=3/2$} \label{Sec:ell-n-choice-3}

When $n=3/2$, it follows from Eq.~\eqref{eq:B_asantz1} jointly with the integral  representation of the underlying hypergeometric function that
\begin{align}\label{eq:new2}
B(r) = &1 -\frac{2 \lambda r_d}{3(\ell-2)(\ell-1) r} \Bigg\{1 \notag \\  &-
\left[{(\ell-1) \left(\frac{r}{r_d}\right)^3+\ell \left(\frac{r}{r_d}\right)^{3/2}+1}\right]\,
\left[1+\left(\frac{r}{r_d}\right)^{3/2}\right]^{-\ell}\Bigg\}\,,
\end{align}
where we assume $\ell >2$ owing to the asymptotic flatness condition, and take  $c_1=0$,  corresponding to a nonsingular de Sitter core. Solution \eqref{eq:new2} describes a class of regular spacetimes with effective ADM mass 
\begin{align}
m_{\rm eff} = \frac{\lambda r_d}{3(\ell-2)(\ell-1)}.
\end{align}
To  the best of our knowledge, this class of solutions is presented here for the first time in the literature.

It is  worth noticing that, upon relaxing the constraint $\ell>2$, one can arrive at   marginally asymptotically flat configurations. For instance, Eq.~\eqref{eq:new2} with $4/3<\ell<2$ gives a metric function that still approaches the Minkowski limit at large $r$, but with a  power-law falloff slower than $1/r$. The borderline case $\ell=2$ yields
\begin{align}
 B(r)=&   1+
\frac{2  \lambda r_d}{3 r}
\!\left\{
\frac{\left(r/r_d\right)^{3/2}}
{1+\left(r/r_d\right)^{3/2}}
-
\ln\!\left[1+\left(r/r_d\right)^{3/2}\right]
\right\},
\end{align}
which displays a characteristic logarithmic term.

\subsubsection{Analytic patterns for solutions with even values of $n$} \label{Sec:ell-n-choice-4}

The models analyzed in the previous sections indicate that the general analytic structure of the metric component $B(r)$  depends sensitively on the particular choice of the coefficients $n$ and $\ell$, a feature that can be traced back to the  hypergeometric function appearing in the general formula \eqref{eq:B_asantz1}.  Apart from the specific parameter combinations  examined before (i.e., $\ell=1+3/n$, $n=3$, and $n=3/2$), we find that, remarkably, regular, strictly asymptotically flat geometries can be organized according to  a systematic  pattern for certain even values of $n$,  depending on  whether $\ell$ is an integer or a half-integer. 

\begin{table}[htp]
\centering
\begin{tabular}{ccc}
\toprule
$n$ & $\ell$ &  Metric function $B(r)$ \\
\midrule

$2$ 
& $\frac{3}{2}$ 
& $\displaystyle 
1 + \frac{\lambda r_d }{\sqrt{r^2 + r_d^2}}
- \frac{\lambda r_d}{r}
\ln\!\left(
\frac{r + \sqrt{r^2 + r_d^2}}{r_d}
\right)$ \\[2ex]

%$2$  & $2$  & $\displaystyle  1+\frac{\lambda r_d }{2} \left[ \frac{r_d}{r^2+r_d^2} - \frac{1}{r}\arctan\!\left(\frac{r}{r_d}\right) \right]$ 

%$2$  & $3$  & $\displaystyle  1 - \frac{\lambda r_d^2  (r^2 - r_d^2)}{8 (r^2 + r_d^2)^2} - \frac{\lambda r_d}{8r} \arctan\!\left(\frac{r}{r_d}\right)$ 

$2$ 
& $\frac{7}{2}$ 
& $\displaystyle 
1
-\frac{\lambda r_d r^2\,(5r_d^2+2r^2)}
{15\,(r^2+r_d^2)^{5/2}}$ \\[2ex]

$2$ 
& $4$ 
& $\displaystyle 
1 -\frac{\lambda r_d^2}{48}\,
\frac{\left(3r^{2}-r_d^{2}\right)\left(r^{2}+3r_d^{2}\right)}
{\left(r^{2}+r_d^{2}\right)^{3}}
-\frac{\lambda r_d}{16r}
\arctan\!\left(\frac{r}{r_d}\right)$ \\[2ex]

$2$ 
& $\frac{9}{2}$ 
& $\displaystyle 
1 -\frac{\lambda r_d r^2 \,
\left(8r^4+28r^2r_d^2+35r_d^4\right)}
{105\,(r^2+r_d^2)^{7/2}}$ \\[2ex]

%$2$  & $5$  & $\displaystyle  1-\frac{\lambda r_d^2\Bigl(15r^6+55r^4r_d^2+73r^2r_d^4-15r_d^6\Bigr)} {384\,(r^2+r_d^2)^4} -\frac{5\lambda r_d}{128r} \arctan\!\left(\frac{r}{r_d}\right)$  \\[2ex]
\midrule %\\[0.2ex]
$4$ 
& $2$ 
& $\displaystyle 
1-\frac{\lambda r_d^2 r^2 }{4(r^4+r_d^4)}-\frac{\lambda r_d}{4 r} I(r)$  \\[2ex]

$4$ 
& $3$ 
& $\displaystyle 
1-\frac{\lambda r_d^2 r^2 (5r^4+9 r_d^4)}{32(r^4+r_d^4)^2}-\frac{5\lambda r_d}{32 r} I(r)$ \\[2ex]
\midrule %\\[0.2ex]
$6$ 
& $1/2$ 
& $\displaystyle 
1-\frac{\lambda r_d}{3r}
\ln\!\left(
\frac{r^{3}}{r_d^{3}}
+\sqrt{1+\frac{r^{6}}{r_d^{6}}}
\right)$ \\[2ex]

$6$ 
& $2$ 
& $\displaystyle 
1
-\frac{\lambda r_d ^4 r^2}{6(r^6+r_d^6)}
-\frac{\lambda r_d}{6r}
\arctan\!\left(\frac{r^3}{r_d^3}\right)$ \\[2ex]

$6$ 
& $\frac{5}{2}$ 
& $\displaystyle 
1-\frac{\lambda r_d r^2\left(2r^6+3r_d^6\right)}
{9\,(r^6+r_d^6)^{3/2}}$ \\[2ex]

$6$ 
& $3$ 
& $\displaystyle 
1-\frac{\lambda r_d^4 r^2 \left(3 r^6+5 r_d^6\right)}{24 \left(r^6+r_d^6\right)^2}-\frac{\lambda r_d}{8r}
\arctan\!\left(\frac{r^3}{r_d^3}\right)$ \\[2ex]

$6$ 
& $\frac{7}{2}$ 
& $\displaystyle 
1-\frac{\lambda r_d r^2\left(8 r^{12}+20 r^6 r_d^6+15 r_d^{12}\right)}
{45 (r^6+r_d^6)^{5/2}}$ \\[2ex]

%$6$  & $4$  & $\displaystyle  1-\frac{\lambda r_d^4 r^2  \left(15 r^{12}+40 r^6 r_d^6+33 r_d^{12}\right)}{144 \left(r^6+r_d^6\right)^3}-\frac{5\lambda r_d}{48r} \arctan\!\left(\frac{r^3}{r_d^3}\right)$ \\[2ex]

\bottomrule
\end{tabular}
\caption{
Representative regular metrics sourced by the generalized power-law density profile \eqref{eq:rho_asantz1}  for $n=2,4,6$ and integer or half-integer values of $\ell$. The solutions display a systematic analytic  pattern. Geometries with $n=2$ and $n=6$ share a similar form, while those with $n=4$ involve the function \( I(r) \)  (see Eq. \eqref{eq:defI}) when $\ell$ is an  integer. Furthermore, configurations satisfying $n\ell=3$ correspond to marginally asymptotically flat spacetimes, as signaled by the  logarithmic correction. Except for the case with $n=2,\ell=3/2$,  previously examined  in Ref.~\cite{Kar:2025phe} (where a minor typo appears in Eq. (3.2)), the remaining solutions appear here for the first time, to the best of our knowledge. }
\label{tab:general_power_law_other}
\end{table}

For example, when $n=2$ and $\ell$ is an integer larger than one (i.e., $\ell=2,3,4,5,\dots$), $B(r)$ can be generically decomposed into the sum of a rational function $\mathscr{R}_{\ell}\left(r^2\right)$ of $r^2$ and a term proportional to $r^{-1}\arctan(r/r_d)$. Schematically, we can write  
\begin{equation}
B(r)=1+\mathscr{R}_{\ell}\left(r^2\right)
+\frac{c_{\ell}}{r}\arctan\!\left(\frac{r}{r_d}\right),
\label{B-r-R-l-form}
\end{equation}
where $c_{\ell}$ is a constant depending on $\ell$. Within this family,   configurations with $n=\ell=2$ and $n=2,\ell=3$ have already been studied in Refs.~\cite{Dymnikova:2004zc} and \cite{Bahamonde:2026bvh}, respectively. In contrast, when $\ell$ is a half-integer of the form $\ell=(2k+1)/2$ with $k\in\mathbb{N}$, the $\arctan$ contribution disappears and the solution reduces to a purely algebraic (fractional) expression involving powers of $r^2$.  The special scenario with $\ell=3/2$ corresponds to a marginally asymptotically flat spacetime (since in this case $n\ell=3$), and thus  contains a logarithmic factor.

An analogous structure emerges for $n=6$. Indeed, an integer  $\ell$ gives rise to a decomposition into a rational function of $r^2$ plus a term proportional to 
$r^{-1}\arctan(r^3/r_d^3)$, whereas a half-integer  $\ell=(2k+1)/2$ leads to  fractional forms comprising powers of $r^2$ and no inverse trigonometric functions.

When $ n = 4 $ and  \( \ell \) is an integer, the metric typically consists of  a fractional expression supplemented by the auxiliary function \( I(r) \) defined in Eq.~\eqref{eq:defI}, which includes both arctangent and logarithmic terms; on the other hand,  for half-integer \( \ell \),  $B(r)$ cannot be written in terms of elementary functions. 

The  analytic structures identified above are illustrated  in Table \ref{tab:general_power_law_other}, where we list  sample regular solutions according to the values of $n$ and $\ell$.

\subsection{Exponentially  suppressed density profiles: $\rho(r)=\rho_0\exp[-(r/r_d)^n]$}
\label{sec:exp1}

The density function with exponential decay
\begin{align}\label{eq:rho_asantz2}
    \rho(r)=\rho_0\exp[-(r/r_d)^n]\,,
\end{align}
is sometimes referred to as  Einasto profile~\cite{Einasto:2009zd,Retana_Montenegro_2012,Konoplya:2025ect} and  has been extensively employed in the investigation of dark matter halos. With $\rho_0>0$ and finite, conditions \eqref{eq:requirerho} hold whenever $n \geq 1$, while the central isotropy constraint \eqref{eq:addtion_req-2} requires $n>1$. On the other hand,  the DEC is always violated in this setting   for $r>r_d(4/n)^{1/n}$.

The metric function corresponding to  the \emph{ansatz} \eqref{eq:rho_asantz2} 
can be cast in terms of the incomplete Gamma function $\Gamma(a,z)$~\cite{abramowitz1968handbook}, namely
\begin{align}\label{eq:B_asantz2}
 B(r) = 1 + \frac{c_1}{r}+\frac{\lambda r_d}{n r} \,
\Gamma \Bigl[3/n, (r/r_d)^n\Bigr] .
\end{align}

In general, it is not possible to formulate $\Gamma(a,z)$ via elementary functions.  However, when $a \equiv 3/n$ is a positive integer or half-integer, such reductions  are possible (cf. Sec.~6.5 of Ref.~\cite{abramowitz1968handbook}), and the relevant cases with $n>1$ are as follow:
\begin{itemize}
\item[-] integer $a$:
\begin{itemize}
\item[$\bullet$] $a=1$ $\Rightarrow$ $n=3$,
\item [$\bullet$] $a=2$ $\Rightarrow$ $n=3/2$,
\end{itemize}
\item [-]  half-integer $a$:
\begin{itemize}
\item [$\bullet$] $a=1/2$ $\Rightarrow$ $n=6$, 
 \item [$\bullet$] $a=3/2$ $\Rightarrow$ $n=2$,
\item [$\bullet$] $a=5/2$ $\Rightarrow$ $n=6/5$.
\end{itemize}
\end{itemize}

The scenarios with  integer $a$ listed above yield purely elementary expressions, while those with half-integer $a$  also involve the error function $\operatorname{erf}(z)= \tfrac{2}{\sqrt{\pi}} \int_0^z \exp\left[-t^2\right] \dd t$ (in particular, the one with $a=3/2$ has been recently  examined in Ref.~\cite{Konoplya:2025ect}). The former will be examined  in Secs. \ref{Sec:n-3-Gamma} and \ref{Sec:n-6-Gamma}.

\subsubsection{Dymnikova black hole ($n=3$)}\label{Sec:n-3-Gamma}

The case $n=3$ (i.e., $a = 1$) is particularly noteworthy. In fact, the identity
\begin{equation}
\Gamma(1,z) = e^{-z},
\end{equation}
with $z \equiv (r/r_d)^3$, gives the general solution 
\begin{equation}\label{eq:Bsol_exp_n3}
B(r)=1+\frac{c_1}{r}+\frac{\lambda r_d}{3r}   \exp\!\left[ -(r/r_d)^3 \right].
\end{equation}
Regularity at $r=0$ sets the integration constant to
$c_1=-\lambda r_d/3$, yielding
\begin{equation}\label{eq:n3}
B(r)=1-\frac{\lambda r_d}{3r}\left\{1-\exp\!\left[ -(r/r_d)^3 \right]\right\},
\end{equation}
which reproduces the well-known Dymnikova nonsingular black hole solution~\cite{Dymnikova1992},  with  effective ADM mass $m_{\rm eff}= \lambda r_d/6$. Remarkably,   our framework allows  this geometry to arise from the unified and general density profile \eqref{eq:rho_asantz2}, rather than from the effective vacuum stress-energy tensor adopted in Ref. \cite{Dymnikova1992}.

\subsubsection{New regular solutions ($n=3/2$)} \label{Sec:n-6-Gamma}

In the scenario with $n=3/2$, we can resort to the result
\begin{equation}
\Gamma\!\left[2, (r/r_d)^{3/2}\right]
=\left[ (r/r_d)^{3/2} + 1 \right]
\exp\!\left[-(r/r_d)^{3/2}\right],
\end{equation}
stemming from the  identity $\Gamma(2,z) = (z+1)e^{-z}$ \cite{abramowitz1968handbook}. In this way, starting from Eq.~\eqref{eq:B_asantz2}, we find that the metric function $B(r)$ admits the closed form
\begin{equation}\label{eq:n32}
B(r)=1-\frac{2\lambda r_d}{3r}
\left\{1-\ee^{-(r/r_d)^{3/2}}\left[(r/r_d)^{3/2}+1\right]\right\},
\end{equation}
where  we have  fixed the integration constant to  $c_1=-2\lambda r_d/3$.  This solution exhibits a de Sitter core and is  strictly asymptotically flat,  with an effective ADM mass $m_{\rm eff}= \lambda r_d/3$. To the best of our knowledge, this  regular spacetime  has never been derived in the existing literature.

\subsection{The composite density profile  $\rho(r)=\rho_0(r/r_d)^{\ell}\exp[-(r/r_d)^n]$} \label{sec:2.5}

The power-law-modulated exponentially suppressed  profile
\begin{equation}\label{eq:anatzexp2}
\rho(r)=\rho_0\left(\frac{r}{r_d}\right)^{\ell}
\exp\!\left[-\left(\frac{r}{r_d}\right)^n\right],
\end{equation}
generalizes the density function analyzed in Sec. \ref{sec:exp1} (see Eq. \eqref{eq:rho_asantz2}). With this parameterization,  the monotonicity assumption \eqref{eq:requirerho-3} furnishes
\begin{align}
\ell \leq 0 \,,\qquad n \geq 0 \,,
\end{align}
while the finiteness requirement of $\rho(r)$ at the origin further enforces $\ell=0$, thereby reducing the model to the class discussed before.

If the energy criteria are relaxed, regular spacetimes may still be obtained for other choices of parameters, with
\begin{align}
B(r)=1+\frac{c_1}{r}+\frac{\lambda r_d }{nr}\,\Gamma\!\left[\frac{l+3}{n},\left(\frac{r}{r_d}\right)^n\right].    
\end{align}
For example, the regular geometries considered in Refs.~\cite{Culetu:2014lca,Lan:2023cvz} can be identified  with the  choice $\ell = -4$ and $n = -1$, while the model studied in Ref.~\cite{Xiang:2013sza} corresponds to $\ell = -5$ and $n = -2$.  However, these solutions generically violate the NEC in the vicinity of the origin and   will not be considered  in this paper.

\subsection{Superposition of regular solutions}
\label{sec:5}

Within the Einstein field equations \eqref{eq:rhoPoutBH}, the relation for the   energy density $\rho(r)$, which can be equivalently  written as (cf. Eq. \eqref{eq:rho_B})
\begin{align}
\rho = \frac{1-B-rB'}{8\pi r^2}\,,
\end{align}
is linear in the metric function $B(r)$ and its derivative. Owing to this   property,   the superposition principle applies, so that any linear combination of the  solutions examined before produces a new regular model whose energy density is given by the corresponding linear combination of the individual density profiles. 

More precisely, let $B_1(r)$ and $B_2(r)$ denote two regular geometries associated with the densities $\rho_1(r)$ and $\rho_2(r)$, respectively. Then
\begin{align}\label{B-r-tot}
B(r) = A_1 B_1(r) + A_2 B_2(r) + 1 - (A_1+A_2),
\end{align}
is also  regular, and is sourced by 
\begin{align}
    \rho(r) = A_1 \rho_1(r) + A_2 \rho_2(r),
\end{align}
with $A_1$ and $A_2$  arbitrary constants; if, in addition, $A_1,A_2\ge0$, the WEC properties of the individual configurations are preserved. The constant terms occurring on the right-hand side of Eq. \eqref{B-r-tot} guarantee that  $B(r)$ attains  the Minkowski form for $r \to \infty$ (see Eq. \eqref{eq:addtion_req}); in addition, if the two metrics are strictly asymptotically flat and admit effective ADM  masses $m_{1,{\rm eff}}$ and $m_{2,{\rm eff}}$, the resulting spacetime remains strictly asymptotically flat, with an effective ADM mass $A_1 m_{1,{\rm eff}}+A_2 m_{2,{\rm eff}}$.

The superposition principle thus enables the systematic construction of new families of regular geometries from simpler building blocks. The generated configurations  inherit key properties  of the original ones, such as   regularity at the center and their  asymptotic behavior. As an  example, by combining the rational falloff profiles with $n=3$ and $n=6$ (see Eq. \eqref{eq:rho_asantz0})
\begin{align}
    \rho(r)= \rho_0 
    \left[
    \frac{1}{1+(r/r_d)^3}
    +
    \frac{1}{1+(r/r_d)^6}
    \right],
\end{align}
we are led to  (cf. Eqs. \eqref{eq:sol_ln0} and \eqref{eq:sol_arctan})
\begin{align}
    B(r)=1 -\frac{\lambda r_d}{3r}\ln\!\left(\frac{r^3}{r_d^3}+1\right)
    -\frac{\lambda r_d}{3r}\arctan\!\left(\frac{r^3}{r_d^3}\right).
\end{align}
In this case, the core energy density becomes $\rho(0)=2\rho_0$,  
and the spacetime is marginally asymptotically flat due to the logarithmic contribution.

\section{Horizons and photon spheres}
\label{sec:4}

In this section we discuss, in a model-independent manner, the conditions for the existence of horizons and photon spheres in the  regular spacetimes constructed in Sec.~\ref{sec:3}.  These structures play a central role in gravitational lensing and black hole shadow phenomenology (see e.g. Refs. \cite{EventHorizonTelescope:2019dse,EventHorizonTelescope:2022wkp,Cardoso-Pani2019,Vagnozzi:2022moj,Vertogradov2024a,Wang-Battista2025,Wang:2025czc,Liu:2025wwqa,Sun:2024xtf,Wei:2026wbg}), and determine whether a given solution represents a regular black hole (featuring  both a horizon and  a photon sphere), a compact object (which lacks a horizon, but possesses at least one photon sphere), or a generic horizonless regular configuration with no photon sphere.  

The general analysis is given in Sec. \ref{Sec:gen-analysis}, while  Sec. \ref{Sec:EHT-analysis} discusses the observational implications of the resulting photon sphere structure in  light of the Event Horizon Telescope (EHT) shadow-size constraints.

\subsection{General analysis}\label{Sec:gen-analysis}

For the static, spherically symmetric metrics of the form~\eqref{eq:metric_static},  horizons are located at the hypersurfaces  $r=r_h$ where their temporal component $g_{tt}$ vanishes, namely
\begin{align}
B(r_h)=0 \,.\label{horizon-identity}
\end{align}
If such a root is admitted and is simple, $B'(r_h)\neq 0$, the surface gravity is finite and the solution describes a non-extremal black hole, while degenerate scenarios satisfying $B(r_h)=B'(r_h)=0$ correspond to extremal cases.

In many regular black hole models, $B(r)$ presents two distinct positive roots, giving rise to an outer (event) horizon and an inner (Cauchy) horizon \cite{Dymnikova2003,Hayward2005}. 
This causal structure is reminiscent of the (non-extremal) Reissner-Nordstr\"om geometry, where the electric    charge likewise leads to a  double-horizon pattern. 
However, spacetimes containing an inner (Cauchy) horizon are generically unstable under linear perturbations~\cite{Wald}, and  in particular the well-known mass-inflation mechanism~\cite{Poisson:1989zz,Ori:1991zz} implies that small perturbations can be infinitely blueshifted near the Cauchy horizon. This phenomenon causes  the  effective mass function to diverge, thereby converting the Cauchy horizon into a mass-inflation singularity~\cite{Ori:1991zz}. Therefore, although the existence of two horizons is a common feature of regular black holes, their physical viability  necessitates a careful stability investigation~\cite{Carballo-Rubio:2018pmi}. This issue lies beyond the scope of this  work, but should be borne in mind when interpreting multi-horizon solutions constructed within our framework.

We now turn to the analysis of circular photon orbits, which define photon spheres. The null geodesic equation in the  equatorial plane $\theta=\pi/2$ of the spacetime \eqref{eq:metric_static}  is given by
\begin{subequations}
\label{null-geo-radial}
  \begin{align}
\dot r^2 + V_{\rm eff}(r)=E^2,
\end{align}
{\rm with}
\begin{align}
V_{\rm eff}(r)= \frac{L^2}{r^2} B(r),
\end{align}
\end{subequations} 
where $E$ and $L$ are the conserved energy and angular momentum, respectively, and an overdot denotes differentiation with respect to the affine parameter. Circular null trajectories correspond to special solutions of the radial dynamics \eqref{null-geo-radial} occurring  at $r=r_{\rm ph}$, where $\dot{r}=0$ and $ \partial {V_{\rm eff}}/\partial r=0$. This yields the geometric identity 
\begin{align}
y(r_{\rm ph}) := 2 B(r_{\rm ph}) - r_{\rm ph} B'(r_{\rm ph})=0\,,
\label{eq:photon_general}
\end{align}
which, via the mass function defined in Eq. \eqref{eq:B_mass}, leads to  
\begin{align}
\frac{r_{\rm ph}}{3}
=m(r_{\rm ph})-m_{\rm loc}(r_{\rm ph}),
\label{eq:photon_mass_relation}
\end{align}
with
\begin{align}
m_{\rm loc}(r_{\rm ph})
\equiv
\frac{4\pi}{3} r_{\rm ph}^3 \rho(r_{\rm ph})\,.
\end{align}

The above quantity can be interpreted as the local effective mass associated with the energy density at the photon sphere radius. It represents  the mass that would be contained inside a sphere of radius $r_{\rm ph}$ if the density were homogeneous and equal to its boundary value $\rho(r_{\rm ph})$.  Relation \eqref{eq:photon_mass_relation} thus shows that the existence of a photon sphere is controlled by the competition between the cumulative mass $m(r_{\rm ph})$ and the local effective term $m_{\rm loc}(r_{\rm ph})$. In the vacuum limit, $\rho(r)=0$ and $m(r)=\mathrm{constant}\equiv M$, and one immediately recovers the standard Schwarzschild result $r_{\rm ph}=3M$. On the other hand, for sufficiently concentrated matter distributions,  $m(r_{\rm ph})$ dominates over $m_{\rm loc}(r_{\rm ph})$, allowing circular null paths to exist even in the absence of horizons. 
In contrast, for more diffuse density profiles, the local effective contribution becomes comparable to the cumulative mass, and Eq.~\eqref{eq:photon_mass_relation} may fail to admit a positive solution. 

This structure highlights an important difference between the conditions defining horizons and photon spheres: while the former depend solely on the function  $2m(r)/r$, which can be related to the compactness of a gravitating  body (conventionally defined as the ratio of its mass to its characteristic radius  \cite{Rosswog2014}),  the latter  probe both the integrated mass and its radial gradient, and are therefore more sensitive to the detailed distribution of matter. 

We also note that an extremal horizon for which $B(r_h)=B'(r_h)=0$ formally solves Eq.~\eqref{eq:photon_general}.  However, this corresponds to a degenerate null geodesic that describes the horizon rather than a genuine photon sphere.

\begin{table}[htp]
\centering
\renewcommand{\arraystretch}{1.2}
\setlength{\tabcolsep}{6pt}
\begin{tabular}{p{1.9cm} p{4.3cm} p{3.8cm} p{3.4cm}}
\toprule
\textbf{{\small Reference}} 
&\textbf{Metric function $B(x)$} 
& \;\;\; \textbf{Horizon} 
& \textbf{Photon sphere} \\
\midrule

{\small Eq.~\eqref{eq:sol_ln0}}& \quad$ 1-\dfrac{\lambda }{3x}\ln\!\left(x^3+1\right)$
& 
\begin{tabular}[t]{@{}l@{}}
$\lambda_h\approx2.66816$ \\
$r_h^{\rm crit} \approx 2.50935\,r_d$ 
\end{tabular}

&
\begin{tabular}[t]{@{}l@{}}
$\lambda_{\rm ph} \approx 2.49566$ \\
$r_{\rm ph}^{\rm crit} \approx 3.58630\,r_d$ 
\end{tabular}
\\
\midrule

{\small Eq.~\eqref{eq:sol_arctan}}&\quad $1-\dfrac{\lambda}{3x}\arctan\!\left(x^3\right)$
& 
\begin{tabular}[t]{@{}l@{}}
$\lambda_h\approx 3.41005$ \\
$r_h^{\rm crit} \approx 1.29517\,r_d$ 
\end{tabular}

&
\begin{tabular}[t]{@{}l@{}}
$\lambda_{\rm ph} \approx 2.89399$ \\
$r_{\rm ph}^{\rm crit} \approx 1.67954\,r_d$ 
\end{tabular}
\\
\midrule

{\small Eq.~\eqref{eq:new2}}&{\small
$\begin{aligned}
&\,1-\frac{2\lambda}{3(\ell-2)(\ell-1) x}\times \\&\Bigg[1- \frac{(\ell-1)x^3
+\ell x^{3/2}+1}
{\left(1+x^{3/2}\right)^{\ell}}
\Bigg]
\end{aligned}$
}
& 
{\small
$\begin{aligned}
\ell=\tfrac{8}{3}&:\,\lambda_h \approx 8.53347
\\&r_h^{\rm crit} \approx 1.89547\,r_d \\
\ell=3&:\,\lambda_h \approx 10.715
\\&r_h^{\rm crit} \approx 1.5874\,r_d \\
\ell=4&:\,\lambda_h \approx 17.8882 
\\&r_h^{\rm crit} \approx 1.10545 \,r_d \\
\ell=5&:\,\lambda_h \approx 25.8547 
\\&r_h^{\rm crit} \approx 0.872071 \,r_d \\
\ell=6&:\,\lambda_h \approx 34.4839 
\\&r_h^{\rm crit} \approx 0.731391 \,r_d \\
\ell=7&:\,\lambda_h \approx 43.6879 
\\&r_h^{\rm crit} \approx 0.636061 \,r_d \\
%\ell=8&,\,\lambda_h \approx 53.4031 
%\\&r_h^{\rm crit} \approx 0.566563 \,r_d \\
\end{aligned}$
}

&
{\small
$\begin{aligned}
\!\!\ell=\tfrac{8}{3}&:\,\lambda_{\rm ph} \approx 7.87271 
\\&r_{\rm ph}^{\rm crit} \approx 2.76737\,r_d \\
\!\!\ell=3&:\,\lambda_{\rm ph} \approx 9.79886 
\\&r_{\rm ph}^{\rm crit} \approx 2.30522\,r_d \\
\!\!\ell=4&:\,\lambda_{\rm ph} \approx 16.0724
\\&r_{\rm ph}^{\rm crit} \approx 1.5874 \,r_d \\
\!\!\ell=5&:\,\lambda_{\rm ph} \approx 22.9836
\\&r_{\rm ph}^{\rm crit} \approx 1.24345 \,r_d \\
\!\!\ell=6&:\,\lambda_{\rm ph} \approx 30.436
\\&r_{\rm ph}^{\rm crit} \approx 1.03782 \,r_d \\
\!\!\ell=7&:\,\lambda_{\rm ph} \approx 38.3623
\\&r_{\rm ph}^{\rm crit} \approx 0.899367 \,r_d \\
%\ell=8&,\,\lambda_{\rm ph} \approx 46.7125
%\\&r_{\rm ph}^{\rm crit} \approx 0.798951 \,r_d \\
\end{aligned}$
}
\\
\midrule

{\small Eq.~\eqref{eq:n32}}&{\small
$1-\frac{2\lambda}{3x}\!
\left[1-\ee^{-x^{3/2}}(x^{3/2}+1)\right]
$
}
& 
\begin{tabular}[t]{@{}l@{}}
$\lambda_h\approx 3.8706$ \\
$r_h^{\rm crit} \approx 1.91771\,r_d$ 
\end{tabular}

&
\begin{tabular}[t]{@{}l@{}}
$\lambda_{\rm ph} \approx 3.29446$ \\
$r_{\rm ph}^{\rm crit} \approx 2.65145\,r_d$ 
\end{tabular}
\\

\midrule

{\small \begin{tabular}[t]{@{}l@{}}
\text{Table~\ref{tab:general_power_law_other}} \\
($n=2$,$\ell=7/2$)
\end{tabular}}
&\qquad $1
-\frac{\lambda x^2 (5+2x^2)}
{15(x^2+1)^{5/2}}$
& 
\begin{tabular}[t]{@{}l@{}}
$\lambda_h\approx 12.1034$ \\
$r_h^{\rm crit} \approx 1.04377\,r_d$ 
\end{tabular}

&
\begin{tabular}[t]{@{}l@{}}
$\lambda_{\rm ph} \approx 10.6195$ \\
$r_{\rm ph}^{\rm crit} \approx 1.45467\,r_d$ 
\end{tabular}
\\

\midrule

{\small \begin{tabular}[t]{@{}l@{}}
\text{Table~\ref{tab:general_power_law_other}} \\
($n=6$,$\ell=5/2$)
\end{tabular}}
&
\begin{tabular}[t]{@{}l@{}} 
$\qquad 1-\frac{\lambda x^2 \left(2x^6+3\right)}
{9(x^6+1)^{3/2}}$
\end{tabular}
& 
\begin{tabular}[t]{@{}l@{}}
$\lambda_h\approx 5.086$ \\
$r_h^{\rm crit} \approx 0.980029\,r_d$ 
\end{tabular}

&
\begin{tabular}[t]{@{}l@{}}
$\lambda_{\rm ph} \approx 4.03397$ \\
$r_{\rm ph}^{\rm crit} \approx 1.2044\,r_d$ 
\end{tabular}
\\

\bottomrule
\end{tabular}
\caption{Horizon and photon sphere structure for selected new regular spacetimes introduced in Sec.~\ref{sec:3}.  For each metric function $B(x)$ (with $x:=r/r_d$), we report the critical  values $\lambda_h$ and $\lambda_{\rm ph}$ of $\lambda := 8\pi \rho_0 r_d^2$, together with the associated critical radii $r_h^{\rm crit}$ and  $r_{\rm ph}^{\rm crit}$. The parameters $\lambda_h$ and $\lambda_{\rm ph}$ are set by the conditions $B(r)=B'(r)=0$ and $y(r)=y'(r)=0$ (cf. Eq. \eqref{eq:photon_general}), respectively, with the former  controlling the emergence of horizons and the latter that of photon spheres. Two horizons ($\lambda>\lambda_h$), an extremal configuration ($\lambda=\lambda_h$)  at $r=r_h^{\rm crit}$, where the two horizons  coincide, or no horizon ($\lambda<\lambda_h$) can occur. Similarly, the spacetime can feature  two photon spheres ($\lambda>\lambda_{\rm ph}$), a degenerate  scenario ($\lambda=\lambda_{\rm ph}$) at $r=r_{\rm ph}^{\rm crit}$,  or no photon sphere  ($\lambda<\lambda_{\rm ph}$). }
\label{tab:global_structure}
\end{table}

In Table~\ref{tab:global_structure}, we report  the horizon and photon sphere pattern of some representative  regular geometries constructed in Sec.~\ref{sec:3}. Different values of $\lambda := 8\pi \rho_0 r_d^2$ amount to distinct spacetime configurations. Denoting  by $\lambda_h$  the critical parameter determined by  $B(r)=B'(r)=0$, one finds that for $\lambda>\lambda_h$ two horizons appear, $\lambda=\lambda_h$ pertains to an extremal solution, while for  $\lambda<\lambda_h$ the spacetime is  horizonless.

Similarly, the existence of photon spheres is governed by the  threshold $\lambda_{\rm ph}$, at which  $y(r)=0$ and $y'(r)=0$ are simultaneously satisfied, with $y$ defined in Eq.~\eqref{eq:photon_general}. Now, for $\lambda>\lambda_{\rm ph}$ two photon spheres are present (in this case, if  horizon(s) exists, the physically relevant photon sphere is the outer one, for which $r_{\rm ph}>r_h$), whereas $\lambda=\lambda_{\rm ph}$ marks the degenerate situation where the two photon spheres merge, and finally no photon sphere exists for $\lambda<\lambda_{\rm ph}$.

A particularly interesting regime arises when $\lambda_{\rm ph}\leq\lambda<\lambda_h$. This parameter window describes regular horizonless geometries  bearing at least one circular null orbit, which typically represents  an outer unstable photon sphere. These features typically qualify compact objects that can potentially mimic several optical properties of black holes  \cite{Cardoso-Pani2019,Bambi2025}, as the presence of a photon sphere leads to strong light bending and can produce  observational signatures such as a shadow-like dark region in gravitational lensing images.

\subsection{Observational tests and  EHT constraints}\label{Sec:EHT-analysis}

\begin{figure}[htbp!]
\centering
\includegraphics[width=0.8\textwidth]{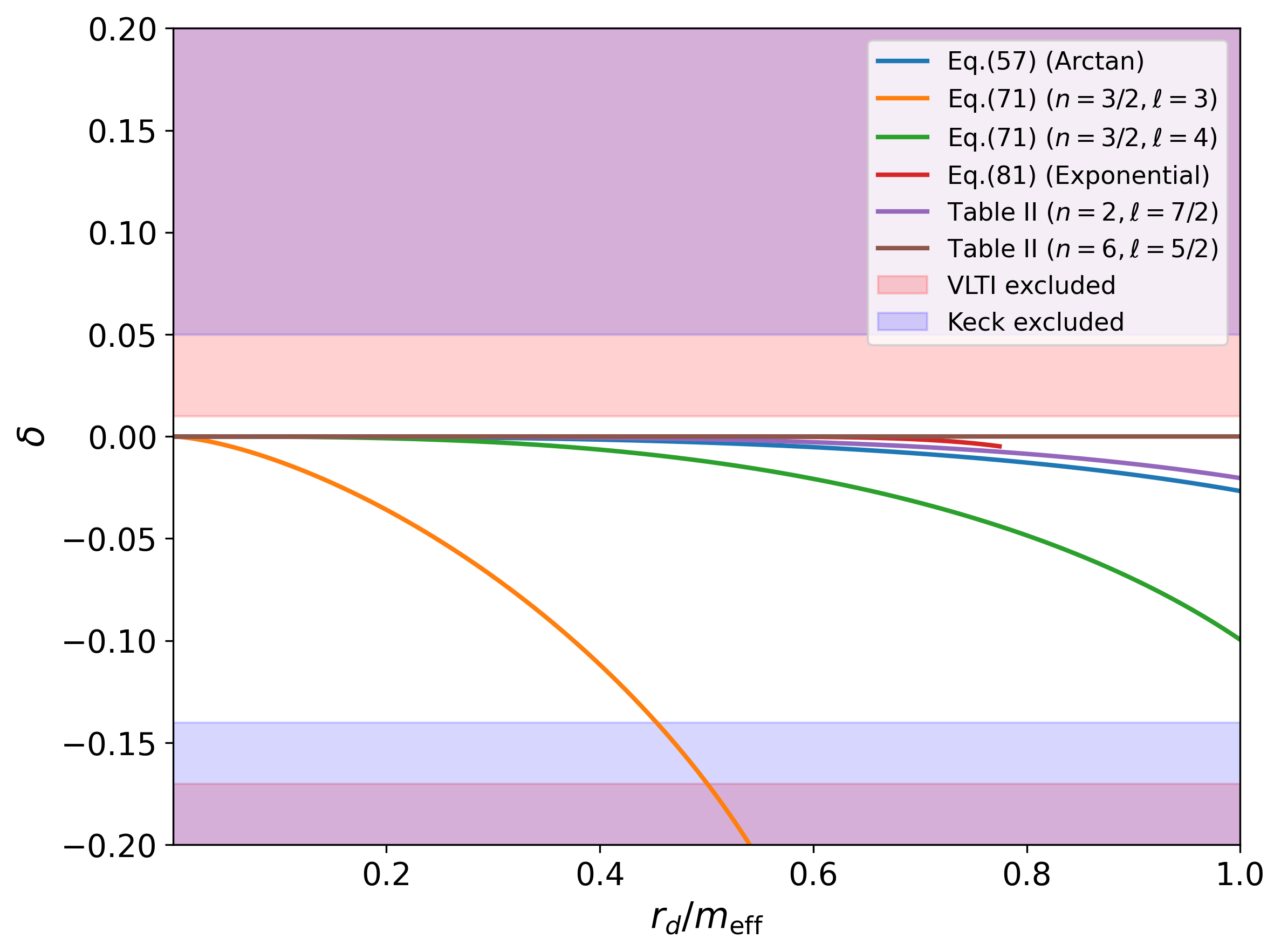}
\caption{Fractional deviation \eqref{eq:delta_b} for some regular black hole candidates as a function of the  effective-ADM-mass-normalized ratio  $r_d/m_{\rm eff}$ (see Eq. \eqref{m-eff-relation}).    The blue and red shaded regions denote the parameter ranges excluded by the Keck- and VLTI-based Sgr~A$^\ast$ constraints \eqref{Keck-1} and \eqref{VLTI-1}, respectively. Only  black hole branches with an outer horizon and an exterior photon sphere are shown. The  exponential solution \eqref{eq:n32} (red curve) is shown only up to the extremal value $r_d/m_{\rm eff}\simeq0.775$, where the inner and outer horizons merge; beyond this point, the geometry becomes horizonless. The smallest deviation from the Schwarzschild limit  $\delta=0$ occurs for the configuration in Table~\ref{tab:general_power_law_other} with $(n,\ell)=(6,5/2)$  (brown curve).} 
\label{fig:EHT-constraint}
\end{figure}

In this section, we briefly discuss the observational relevance of the photon sphere analysis carried out before. Among the possible phenomenological probes, we focus on EHT black hole shadow measurements. A full comparison with the EHT images requires a radiative-transfer model for the accretion flow and a visibility-domain study~\cite{EventHorizonTelescope:2019dse,EventHorizonTelescope:2022xqj}, which are beyond the scope of this paper. Nevertheless, we can perform a simple geometric consistency check using the critical impact parameter associated with the unstable photon sphere, which for  metrics \eqref{eq:metric_static}  is given by~\cite{Li:2026amn,Vagnozzi:2022moj}
\begin{equation} 
b_c = \frac{r_{\rm ph}}{\sqrt{B(r_{\rm ph})}}\,, \label{eq:bc-photon-sphere} 
\end{equation} 
where  $r_{\rm ph}$ can be obtained from Eq.~\eqref{eq:photon_general}. 

Following the examination of
Refs.~\cite{EventHorizonTelescope:2022xqj,Vagnozzi:2022moj}, we focus on Sgr~A$^\ast$, whose accurately
measured mass-to-distance ratio allows the observed angular ring size to be directly translated into bounds on the dimensionless shadow deviation from the Schwarzschild prediction. This enables a
model-independent comparison through the fractional deviation parameter
\begin{equation} 
\delta := \frac{b_c}{b_c^{\rm S}}-1, \label{eq:delta_b} 
\end{equation} 
with $b_c^{\rm S}=3\sqrt{3}M$ the standard Schwarzschild value. The choice of Sgr~A$^\ast$ is motivated by the following two reasons. As emphasized in Ref.~\cite{Vagnozzi:2022moj}, Sgr~A$^\ast$ samples a much lower mass strong-curvature regime than M87$^\ast$, which can lead to stronger constraints on dimensionful departures from general relativity. More importantly for the present study, the mass-to-distance ratio of Sgr~A$^\ast$ is known with greater accuracy than that of M87$^\ast$, enabling a more direct determination of the dimensionless shadow deviation. 
For Sgr~A$^\ast$,  the EHT collaboration reports~\cite{EventHorizonTelescope:2022xqj} 
\begin{equation} \delta = -0.04^{+0.09}_{-0.10}, \qquad \text{(Keck prior)}, \label{Keck-1}
\end{equation} and 
\begin{equation} \delta = -0.08^{+0.09}_{-0.09}, \qquad \text{(VLTI prior)}. \label{VLTI-1}
\end{equation} 
These  provide a first empirical  test of the regular models derived in this paper, as illustrated in   Fig.~\ref{fig:EHT-constraint}. Among the configurations considered, the rational solution displayed in Table~\ref{tab:general_power_law_other} with
$(n,\ell)=(6,5/2)$ remains the closest  to the Schwarzschild case (corresponding to $\delta=0$) over its black-hole branch (as expected,  because its large-radius correction to the Schwarzschild term starts only at order $\mathcal O(m_{\rm eff}\, r_d^{12}/r^{13})$).

\section{Rotating solutions via the Newman-Janis algorithm}\label{Sec:NJ-rotating}

In this section, we extend our analysis to rotating models by resorting to the Newman-Janis algorithm. We first apply the method to the  families of regular static spacetimes studied in Sec. \ref{sec:3}, analyzing their main properties and regularity (Sec. \ref{Sec:NJ-applicatios}). We then focus on some representative  configurations and examine their shadow profiles in Sec. \ref{Sec:shadow-rotating}.

We assume that the reader is acquainted with the Newman-Janis formalism and refer e.g.  to  Refs. \cite{Drake1998,Erbin2014} for further details.

\subsection{Properties of  rotating configurations}\label{Sec:NJ-applicatios}

The Newman-Janis procedure  provides a recipe to generate stationary, axisymmetric geometries from a spherically symmetric seed metric by means of  a particular complexification procedure. When  applying it to the Kerr-Schild solutions worked out before, two  points deserve particular attention. First, the  complexification  is  inherently   ambiguous, and one must ensure that the ensuing rotating spacetime admits a well-defined transformation to Boyer-Lindquist coordinates. Second,  the  approach does not, in general, preserve the matter model supporting the seed solution; thus,  it is necessary to verify whether the final configuration can be interpreted within  the same matter framework as the original one. 

Bearing these premises in mind, a convenient  way to proceed  is to adopt the so-called type-I   prescription (see e.g. Ref. \cite{Bambi:2013ufa}; for other applications see e.g. Refs. \cite{Brahma:2020eos,Xu:2021lff,Mazza:2021rgq,Ali:2022yrf,Kamenshchik2023b,Pantig:2026bhx}). Recalling that for our regular models $B(r)$  is given by Eq. \eqref{eq:B_mass},  in the type-I scheme  the factor $1/r$ entering $B(r)=1-2m(r)/r$ is complexified  \qm{\emph{\`a la} Schwarzschild} via the  replacement rule \cite{Erbin2014}
\begin{align}
    \frac{1}{r}  \rightarrow \frac{1}{2} \left(\frac{1}{r}+\frac{1}{\bar{r}}\right),
\end{align}
the overbar denoting  complex conjugation,  while the Misner-Sharp-Hernandez mass $m(r)$ is  left unchanged. With this choice, the transformation to Boyer-Lindquist coordinates $\mathscr{X}^\alpha=\left(t,r,\theta,\phi\right)$ is integrable, and the  resulting rotating metric $\mathscr{G}_{\alpha \beta}$ attains the Kerr-like form  
\begin{align}
\dd s^2 =\mathscr{G}_{\alpha \beta} \, \dd \mathscr{X}^\alpha \dd \mathscr{X}^\beta= -  \dd t^2+ \Sigma \left(\frac{\dd r^2}{\Delta}  +  \dd \theta^2 \right) + \left(a^2  + r^2\right) \sin^2 \theta \, \dd \phi^2 + \frac{2r m(r)}{\Sigma}\left( \dd t -a \sin^2 \theta \, \dd \phi\right)^2,    \label{rotating-metric-general}
\end{align}
with
\begin{align}
\Sigma:=&r^2 + a^2 \cos^2 \theta,
 \nonumber \\
\Delta:=& r^2 + a^2 - 2r m(r), \label{eq:Delta}
\end{align}
and $a$ representing the rotation parameter. The (Petrov type-D) spacetime \eqref{rotating-metric-general} belongs to the so-called G\"{u}rses-G\"{u}rsey class \cite{Gurses1975,Maeda:2021jdc}, and has a Hawking-Ellis type-I \cite{Hawking-Ellis1973} stress-energy tensor. Accordingly, it can be diagonalized in an appropriate local orthonormal frame, where it retains the same anisotropic-fluid algebraic structure as in the Kerr-Schild case. The associated  energy density and principal pressures are given by  (see e.g. Refs. \cite{Burinskii:2001bq,Maeda:2021jdc} for  further details)
\begin{align}
\rho&=-p_1=\frac{r^2 m'}{4 \pi \Sigma^2},
\nonumber \\
p_2&=p_3=-\frac{r\Sigma m''+2m' a^2 \cos^2 \theta}{8 \pi\Sigma^2}, \label{stress-energy-rotating}
\end{align}
and reduce to the corresponding static  formula \eqref{eq:rhoPoutBH} in the limit $a\to0$. We emphasize that the above quantities refer to the chosen local orthonormal frame, as the stress-energy tensor is not, in general, diagonal in the Boyer-Lindquist coordinate basis.

The above relations allow us to assess whether the matter content sustaining solution \eqref{rotating-metric-general} complies with classical energy conditions. It is clear from Eq. \eqref{stress-energy-rotating} that, while the radial NEC is saturated identically (i.e.,  $\rho+p_1=0 $), nontrivial constraints come from the tangential directions, where
\begin{equation}\label{stress-energy-rotating2}
\rho+p_2 =
\frac{2m'(r^2-a^2\cos^2\theta)-r\Sigma m''}{8\pi\Sigma^2}.
\end{equation}
Owing to the angular dependence introduced by the spin parameter, the density becomes a two-dimensional function, $\rho=\rho(r,\theta)$. Consequently, the fulfillment of energy criteria cannot be  inferred simply from the static seed solution and must be analyzed independently for each rotating configuration. As an illustration, consider a mass function  admitting around  $r=0$ the expansion
\begin{equation}
m(r)=c_1 r^3+\mathcal{O}(r^4), \label{mass-exp-rot}
\end{equation}
which includes the behavior of the static regular scenarios examined before  (see Eqs.~\eqref{eq:m0} and \eqref{density-parametrization}). In this setup,  Eq.~\eqref{stress-energy-rotating2} gives, at leading order,
\begin{equation}
\rho+p_2=-\frac{3c_1a^2r^2\cos^2\theta}
{2\pi\Sigma^2}
+\mathcal{O}(r^3). 
\end{equation}
For $c_1>0$ and  $a \neq 0$, the above combination  is negative away from the equatorial plane (i.e.,  for $\theta\neq \pi/2$),  thereby indicating a violation of the tangential NEC. This means that, although the original Kerr-Schild geometries fulfill the WEC, their rotating Kerr-like counterparts fail to satisfy even the NEC and hence  all standard energy criteria. This result is consistent with the outcome of e.g. Ref. \cite{Maeda:2021jdc}, where  energy conditions are investigated in detail for well-known rotating black holes such as the Bardeen and Hayward models.

Let us now address the question of curvature regularity. The Ricci, Ricci square, and Kretschmann scalars associated with the metric \eqref{rotating-metric-general} read as, respectively,
\begin{align}
R
=&\,
\frac{2\left(rm''+2m'\right)}{\Sigma},
\\
R_{\mu \nu} R^{\mu \nu}
=&\,\frac{2}{\Sigma^4}\left\{
4r^4\left(m'\right)^2+\left[\Sigma\left(rm''+2m'\right)-2r^2m'\right]^2
\right\},
\\
R_{\mu \nu \alpha \beta} R^{\mu \nu \alpha \beta}
=&\,\frac{4}{\Sigma^6}\Bigg\{\Sigma^4\left[
r^2 \left(m''\right)^2+4r m'm''+4\left(m'\right)^2
\right]
\nonumber\\
&\,
-\Sigma^3\left[
12m^2+48rmm'+12r^2mm''+28r^2\left(m'\right)^2+8r^3m'm''\right]
\nonumber\\
&\,
+\Sigma^2\left[216r^2m^2+224r^3mm'+16r^4mm''+32r^4\left(m'\right)^2\right]
\nonumber\\
&\,-\Sigma\left[
576r^4m^2+192r^5mm'\right]+384r^6m^2\Bigg\}.
\end{align}
As proved in Ref.  \cite{Maeda:2021jdc},  the usual Kerr ring curvature singularity at $(r,\theta)=(0,\pi/2)$ is replaced by a ring-like conical singularity   whenever the mass function admits the near-origin expansion $m(r)\sim r^\alpha$ with $\alpha\geq3$, as is the case for all static configurations considered in this paper.

Nevertheless, the finiteness of curvature invariants  does not necessarily imply that the ring-like region is  regular.  This subtlety can be clarified by considering again the situation \eqref{mass-exp-rot}. Neglecting higher-order contributions, we find,  along the equatorial plane
$\theta=\pi/2$, 
\begin{align}
\lim_{r\rightarrow0}R
&=24c_1, \\
\lim_{r\rightarrow0}R_{\mu\nu}R^{\mu\nu}
&=144c_1^2, \\
\lim_{r\rightarrow0}
R_{\mu\nu\alpha\beta}R^{\mu\nu\alpha\beta}
&=96c_1^2 .
\end{align}
On the other hand, approaching the same point along any non-equatorial path \(\theta\neq\pi/2\), the quantity \(\Sigma\) remains finite and nonzero, and one obtains
\begin{equation}
\lim_{r\rightarrow0}R
= \lim_{r\rightarrow0}R_{\mu\nu}R^{\mu\nu}
= \lim_{r\rightarrow0} R_{\mu\nu\alpha\beta}R^{\mu\nu\alpha\beta}
=0 .
\end{equation}
Therefore, the limiting  values of curvature scalars are direction-dependent near the Kerr-like ring  \cite{Bambi:2013ufa}. This feature indicates that the absence of curvature divergences alone is insufficient to establish the complete regularity of  rotating geometries at $\Sigma=0$. This feature stands in stark contrast to the static models. 

\subsection{Analysis of shadow profiles} \label{Sec:shadow-rotating}

In this section, we further analyze the properties of G\"{u}rses-G\"{u}rsey   spacetimes \eqref{rotating-metric-general} by examining the shadow profile of four selected examples. Specifically, we consider the rotating counterparts of the static arctan-modified solution \eqref{eq:sol_arctan}, the mixed arctangent-logarithmic geometry \eqref{B-and-I}, and two additional models in Table~\ref{tab:general_power_law_other}, corresponding to $(n,\ell)=(4,2)$ and $(6,7/2)$.
\begin{figure}[htbp!]
\centering
\includegraphics[width=1\textwidth]{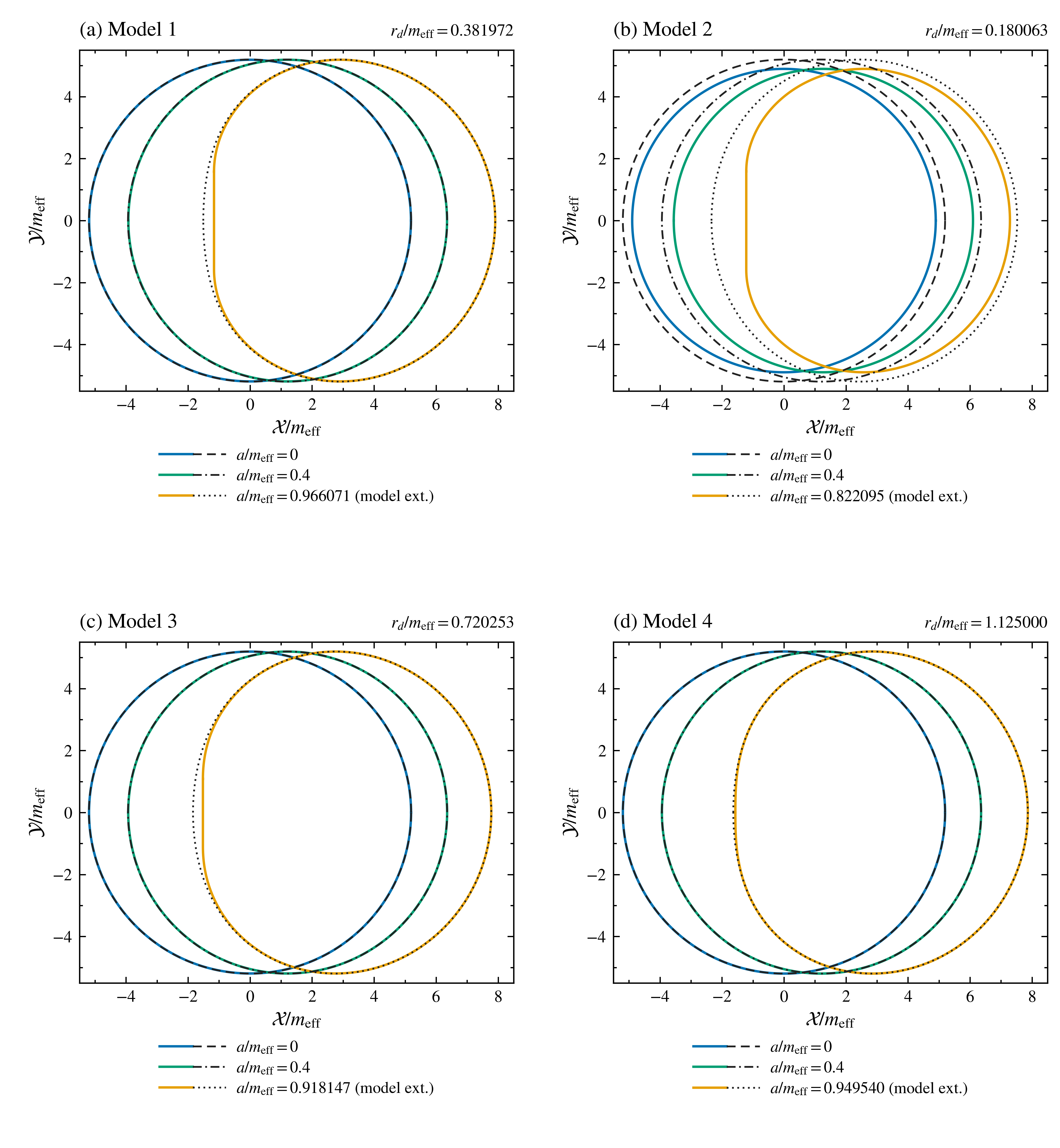}
\caption{Shadow profiles  of models 1-4 given in Eq. \eqref{model-shadow} for a  distant equatorial observer. All configurations have the same effective ADM mass
$m_{\mathrm{eff}}$ and parameter $\lambda=10$, with
$r_d/m_{\mathrm{eff}}\simeq0.381972$, $0.180063$, $0.720253$, and
$1.125000$, respectively.
Colored solid curves correspond to
$a/m_{\mathrm{eff}}=0$, $0.4$, and the respective extremal values
$a_{\mathrm{ext}}/m_{\mathrm{eff}}\simeq
0.966071$, $0.822095$, $0.918147$, and $0.949540$.
Black dashed curves show Kerr black hole shadows with the same mass and spin, with the extremal case, $a/m_{\mathrm{eff}}=1$, not shown. Model 2 deviates the most from the corresponding Kerr shadow. }
\label{fig:shadows-meff-lambda}
\end{figure}
In view of Eq. \eqref{eq:B_mass}, the mass functions $m(r)$   read as, respectively, 
\begin{align}
m_{\rm model\,1}(r) &= \frac{\lambda r_d}{6} \arctan\left(\frac{r^3}{r_d^3}\right)\,,
   \qquad \qquad m_{\rm model\,2}(r) = \frac{\lambda r_d}{2}\,I(r)\,,
   \nonumber \\[2mm]
m_{\rm model\,3}(r) &= \frac{\lambda r_d^2 r^3}{8\left(r^4+r_d^4\right)}+\frac{\lambda r_d}{8}\,I(r)\,,
\qquad
m_{\rm model\,4}(r) = \frac{\lambda r_d r^3 \left(8r^{12}+20r^6r_d^6+15r_d^{12}\right)}{90\left(r^6+r_d^6\right)^{5/2}}\,, \label{model-shadow}
\end{align}
where, we recall, $I(r)$ is given by Eq.~\eqref{eq:defI}. 

The shadow boundary is determined by unstable spherical photon orbits outside the event horizon, satisfying $\mathcal{R}=\partial_r\mathcal{R}=0$ \cite{Tsukamoto:2017fxq}, where the (energy-rescaled) radial potential $\mathcal{R}(r)$ for null geodesic motion  is defined by
\begin{equation}
\mathcal{R}(r)=\left(r^2+a^2-a\xi\right)^2-\kappa \Delta\,,
\qquad
\xi=\frac{\mathcal{L}_z}{\mathcal{E}},\qquad
\kappa=\frac{\mathcal{Q}}{\mathcal{E}^2}+(\xi-a)^2 ,
\end{equation}
with $\mathcal{E}$, $\mathcal{L}_z$, and $\mathcal{Q}$  the photon energy, axial angular momentum, and Carter constant, respectively.

In Fig.~\ref{fig:shadows-meff-lambda}, we compare the shadows of the four models \eqref{model-shadow} at a common effective ADM mass $m_{\rm eff}$ and dimensionless parameter $\lambda=10$ \footnote{The parameters $r_d$, $\lambda$, and $m_{\rm eff}$ satisfy a model-dependent relation and are therefore not independent; see, e.g., Eq.~\eqref{m-eff}. Fixing $\lambda$ and a common $m_{\rm eff}$ uniquely determines $r_d/m_{\rm eff}$ for each model.}, and adopt screen coordinates, which, for a distant observer with inclination $\theta_O$,  are \cite{Perlick:2021aok}
\begin{equation}
\mathcal{X}=a\sin\theta_O-\frac{\xi}{\sin\theta_O},
\qquad
\mathcal{Y}=\pm\sqrt{\kappa-
\left(\frac{\xi}{\sin\theta_O}-a\sin\theta_O\right)^2}.
\end{equation}
All lengths are normalized by the same $m_{\mathrm{eff}}$, and an equatorial observer, located at $\theta_O=\pi/2$, is adopted. The mass normalization fixes $r_d/m_{\mathrm{eff}}=12/(\pi\lambda)$, $4\sqrt{2}/(\pi\lambda)$, $16\sqrt{2}/(\pi\lambda)$, and $45/(4\lambda)$ for metrics 1-4, respectively.  Each panel includes $a/m_{\mathrm{eff}}=0$, $0.4$, and the  model-dependent extremal spin $a_{\mathrm{ext}}$, which is obtained by the conditions $\Delta(r_{\mathrm{ext}})= \Delta'(r_{\mathrm{ext}})=0$.  Among the four models analyzed,  configuration 2 exhibits the largest difference with respect to  the  Kerr black hole scenario, while the fourth one remains relatively close to Kerr framework in size but develops a qualitatively different extremal structure. Solutions 1 and 3 show rather similar shadow profiles and remain close to the corresponding Kerr contours over the spin range considered, with more visible deviations emerging near extremality.  In particular, the extremal critical curves of geometries 1-3 contain a visible horizon-limit vertical segment associated with critical null geodesics approaching the degenerate horizon. For scenario 4, this horizon-limit branch is not visible to the distant observer, and the extremal shadow boundary is instead generated entirely by a detached family of unstable spherical photon orbits outside the horizon.

\section{Concluding remarks}
\label{sec:conclusions}

In this paper, we have devised a method for constructing families of static, spherically symmetric regular spacetimes in general relativity satisfying the WEC.  

The starting point of our construction is given by  the WEC requirements \eqref{eq:requirementwec-1} and \eqref{eq:requirementwec-2}, jointly with the regularity assumption \eqref{eq:requirementwec-3}, the latter    only calling for the boundedness of a single invariant, namely the Kretschmann scalar. Remarkably,  this  condition guarantees the finiteness of all curvature invariants, including the  full set of Zakhary-McIntosh  scalars, and  is  equivalent to causal geodesic completeness  for   all the Kerr-Schild configurations studied in this paper, which display   asymptotically Minkowskian behavior and regular cores. Hypotheses \eqref{eq:requirementwec} translate into the  system of ordinary differential equations \eqref{NEC-WEC-rel1}--\eqref{NEC-WEC-rel4},  supplemented by  the admissibility  criteria \eqref{eq:requirerho} and \eqref{eq:addition_all}  encoding physically reasonable regularity, boundary, and monotonicity conditions on the matter distribution and the metric function $B(r)$. In this way,  regular WEC-obeying solutions  can be immediately obtained by solving Eq. \eqref{eq:rho_B}, which relates $B(r)$ and the energy density $\rho(r)$ via  a linear differential equation (see Sec.~\ref{sec:2}).

The  validity of the WEC provides a physical basis for the spacetimes derived within our scheme, which we have studied in detail in Sec.~\ref{sec:3} by classifying energy-density profiles in terms of their complexity parameter $\mathcal{C}$. Well-known regular models, such as the Bardeen, Hayward, and Dymnikova geometries, or prominent dark-matter halo paradigms,  emerge as particular realizations of a broader and unified framework, where  closed-form expressions for   new regular solutions can also be computed (see, e.g., Eqs. \eqref{eq:sol_arctan},  \eqref{eq:new2}, and \eqref{eq:n32}, and Table \ref{tab:general_power_law_other}). These  may represent, depending on the presence of horizons and photon spheres,  geometric candidates for either black holes or generic compact objects, as explored in Sec.~\ref{sec:4}, where we also performed a preliminary phenomenological assessment based on EHT shadow constraints. In addition, the possibility of introducing rotation has  been taken into account in Sec. \ref{Sec:NJ-rotating} by means of the Newman-Janis algorithm. Despite the fulfillment of the WEC by the static seed geometries, rotation generically spoils the tangential NEC near the center and away from the equatorial plane. Furthermore, the   finiteness of curvature scalars does not translate into the  the complete regularity of the rotating models, owing to their direction-dependent limiting values at the Kerr-like ring. 

However, key aspects of the  new solutions,  such as stability and formation scenarios, have not been addressed, and hence their full astrophysical relevance  remains to be established (for instance, related dynamical aspects of regular black hole formation have recently been explored in  Ref. \cite{Battista2026d}). Another issue deserving further investigation concerns the matching of the regular configurations constructed in Sec.~\ref{sec:3}  to a Schwarzschild exterior. Since their mass function $m(r)$  is monotonically increasing at any finite radius, it is not difficult to prove that a smooth matching is impossible within our framework, and hence  the appearance of a thin shell of matter is unavoidable. In particular, the resulting surface layer is  endowed with  a  stress-energy tensor of diagonal perfect-fluid form, with vanishing  surface energy density and positive tangential pressure. In this setup, the DEC is spoiled, while the remaining energy criteria hold. These results motivate a  systematic study of the junction conditions and of the physical properties of the associated surface layer in future work.

From a broader perspective, our analysis has involved energy densities $\rho(r)$ with a relatively simple analytic structure, as reflected by the fact that they are characterized by  $\mathcal{C}\leq 8$.  If no further profiles with lower complexity  can be identified, it is plausible to conclude that we have explored the simplest choices for $\rho(r)$ that lead to regular spacetimes adhering to the WEC. Nevertheless,  our systematic  approach can  be extended to   general  models  with  more complicated structures. In addition, it allows us to study a broader class of nonsingular  configurations, including, for example, wormholes~\cite{Bouhmadi-Lopez:2021zwt,Battista2024-WH,Mehdizadeh2024,Xu:2025jad,Konoplya:2025hgp,Li:2026mam},  gravastar-like structures~\cite{Jampolski2025a,Chen:2025ywj,Sakib:2026ngq,Rahaman:2026ymb}, or generic exotic compact objects \cite{Cardoso-Pani2019}. Although such extensions have not been explicitly realized here, it is likely that they  could avoid singular behavior while satisfying relevant energy conditions through suitably chosen matter distributions or regularization schemes. Assessing the viability and physical consistency of such hypothetical scenarios constitutes an interesting avenue for future work.

In this regard, a subtlety concerning regularity should be mentioned. In Ref.~\cite{Maeda:2021jdc}, seven criteria have been proposed to determine physically reasonable nonsingular black holes. Configurations such as Bardeen, Hayward, Dymnikova,  Fan-Wang, and their rotating counterparts were found not to fulfill all of them simultaneously (see, e.g., Table 5 in Ref. \cite{Maeda:2021jdc}). This highlights that the study of regular metrics represents an active area of research that merits further analysis.

\section*{Acknowledgments}

The work of Z. W. is supported by the National Natural Science Foundation of China  (12405063). E. B. acknowledges the support of INFN  {\it iniziativa specifica}  Moonlight2. We thank the anonymous referee for the useful comments provided.

\appendix

\section{Riemann tensor components and Zakhary-McIntosh  invariants}\label{Appendix-0}

In this Appendix, we compute the components of the Riemann tensor and the full set of the Zakhary-McIntosh (ZM)  invariants for the static, spherically symmetric  geometry~\eqref{eq:metric_static}.
These computations are most conveniently carried out in an orthonormal basis~\cite{Wald}, using the (inverse) tetrad field defined in Eq.~\eqref{eq:orthonormal_frame}.

The torsion-free condition,  also known as the first Cartan structure equation~\cite{Wald}, can be  written in the language of  differential forms as~\cite{Nakahara2003,Carroll2004}
\begin{align}\label{eq:torsionfree}
    \dd e^a + \omega^{a}{}_{ b} \,\wedge \, e^b=0, 
\end{align}
where $e^a=e^a{}_{\mu} \dd x^{\mu}$ denotes the orthonormal coframe,  $\omega^{a}{}_{b}$  the spin connection one-form, and frame indices $a,b,\dots=\hat{0}, \hat{1}, \hat{2}, \hat{3}$.
For the metric~\eqref{eq:metric_static}, the nonvanishing components of the spin connection are then found to be 
\begin{subequations}\label{eq:spinconnection}
\begin{align}
    \omega^{\hat 0}{}_{ \hat 1} &=-\omega^{\hat 1}{}_{\hat 0}=\frac{B'}{2}\dd t =\frac{B'}{2\sqrt{B}} e^{\hat 0}\,,\\
    \omega^{\hat 1}{}_{\hat 2} &=-\omega^{\hat 2}{}_{\hat  1}=-\sqrt{B} \,\dd \theta =-\frac{\sqrt{B}}{r} e^{\hat 2} \,,\\
    \omega^{\hat 1}{}_{\hat 3} &=-\omega^{\hat 3}{}_{\hat 1}=-\sqrt{B} \sin \theta \,\dd \phi = -\frac{\sqrt{B}}{r} e^{\hat 3} \,,\\
    \omega^{\hat 2}{}_{\hat 3} &=-\omega^{\hat 3}{}_{\hat 2}=-\cos \theta \,\dd \phi =-\frac{\cot \theta}{r} e^{\hat 3}\,,
\end{align}
\end{subequations}
where we recall that a prime denotes the  differentiation with respect to the radial variable $r$. 
The curvature two-form can be obtained from the second Cartan structure equation
\begin{align}\label{eq:Riemann}
R^a{}_{ b}=\dd \omega^{a}{}_{ b}+ \omega^{a}{}_{ c} \,\wedge \, \omega^{c}{}_{ b}\,.
\end{align}
Then, employing Eq.~\eqref{eq:spinconnection}, we obtain the following nonvanishing components of $R^a{}_{ b}$: 
\begin{subequations}
\begin{align}
     R^{\hat 0}{}_{\hat 1}&=- R^{\hat 1}{}_{\hat 0}=\frac{B''}{2} \dd r  \,\wedge \, \dd t = \frac{B''}{2} e^{\hat 1} \wedge e^{\hat 0}\,,\\
     R^{\hat 0}{}_{\hat 2}&=- R^{\hat 2}{}_{\hat 0}=-\frac{\sqrt{B}B'}{2} \dd t  \,\wedge \, \dd \theta =-\frac{B'}{2r} e^{\hat 0} \wedge e^{\hat 2}\,, \\
     R^{\hat 0}{}_{\hat 3}&=- R^{\hat 3}{}_{\hat 0}=-\frac{\sqrt{B}B' \sin \theta}{2} \dd t  \,\wedge \, \dd \phi = -\frac{B'}{2r} e^{\hat 0} \wedge e^{\hat 3}\,,\\  
     R^{\hat 1}{}_{\hat 2}&=- R^{\hat 2}{}_{\hat 1}=-\frac{B'}{2\sqrt{B}} \dd r  \,\wedge \, \dd \theta =-\frac{B'}{2r} e^{\hat 1} \wedge e^{\hat 2} \,,\\   
     R^{\hat 1}{}_{\hat 3}&=- R^{\hat 3}{}_{\hat 1}=-\frac{B'\sin \theta}{2\sqrt{B}} \dd r  \,\wedge \, \dd \phi =-\frac{B'}{2r} e^{\hat 1} \wedge e^{\hat 3} \,,\\  
     R^{\hat 2}{}_{\hat 3}&=- R^{\hat 3}{}_{\hat 2}=(1-B) \sin \theta \,\dd \theta  \,\wedge \, \dd \phi = \frac{1-B}{r^2} e^{\hat 2} \wedge e^{\hat 3}\,.     
\end{align}  
\end{subequations}
It is thus evident that the curvature two-form involves only three independent  functions of $r$, namely (cf. Eq. \eqref{X-Y-Z-text})
\begin{align}\label{eq:XYZ}
    X(r):=\frac{B''}{2} \,, \quad Y(r):=\frac{B'}{2r} \,, \quad Z(r):=\frac{1-B}{r^2}\,.
\end{align}
Now, from the definition of the curvature two-form~\cite{Nakahara2003}
\begin{align}
    R^a{}_{ b} := \frac{1}{2} R^{ a}{}_{b c d}\,
    e^{c} \wedge e^{ d} \,,
\end{align}
one can directly read off the following independent nonvanishing components of the Riemann tensor in the orthonormal frame:
\begin{subequations}
\begin{align}
   R_{\hat{0}\hat{1}\hat{0}\hat{1}} &=X\,,\\
   R_{\hat{0}\hat{2}\hat{0}\hat{2}} &=R_{\hat{0}\hat{3}\hat{0}\hat{3}}=Y\,,\\
   R_{\hat{1}\hat{2}\hat{1}\hat{2}} &=R_{\hat{1}\hat{3}\hat{1}\hat{3}}=-Y\,,\\
    R_{\hat{2}\hat{3}\hat{2}\hat{3}} &=Z\,,
\end{align}  
\end{subequations}
all the other nonvanishing components following from the usual symmetries
\begin{align}\label{eq:Rsymmetry}
    R_{abcd} &= - R_{bacd} = - R_{abdc} = R_{cdab} \,.
\end{align}

Because the frame indices are raised and lowered with the Minkowski metric $\eta_{ab} = \mathrm{diag}(-1,1,1,1)$, the full set of Riemann tensor components--irrespective of index positions--is algebraically determined by the three functions \eqref{eq:XYZ}. Consequently, \emph{all}  algebraic curvature invariants constructed by complete contractions of Riemann tensor, such as the Kretschmann and Ricci scalars, or the  other higher-order ZM invariants, can be expressed  as polynomials in $X(r),Y(r),$ and $Z(r)$ with constant coefficients. Although this result is derived in an orthonormal frame, it is completely general, since curvature  scalars are   independent of the choice of the coordinates.

Therefore, the condition that the three basic curvature functions $X, Y$, and $Z$ remain finite, is both necessary and sufficient to guarantee that \emph{all} algebraic curvature invariants of the spacetime are finite. Thus, considering in particular the center $r=0$ and assuming that $B(r)$ admits the power-series expansion 
\begin{equation}\label{eq:APPB_expansion}
B(r) = B_0 + B_1 r + B_2 r^2 + \mathcal{O}(r^3)\,,
\end{equation}
it follows from Eq.~\eqref{eq:XYZ} that the finiteness of $X$, $Y$, and $Z$ as $r \to 0$
requires
\begin{equation}\label{eq:AAB01}
B_0 = 1\,, \qquad B_1 = 0\,,
\end{equation}
as we have already shown in Eq. \eqref{eq:B01}. 
Conversely, if Eq.~\eqref{eq:AAB01} holds, then $X$, $Y$, and $Z$ remain finite at the origin.
Therefore, Eq.~\eqref{eq:AAB01} provides a necessary and sufficient condition for the boundedness of all algebraic curvature invariants at $r=0$.

For completeness, we now derive the whole set of the seventeen ZM invariants \cite{Eichhorn2021,universe6020022,Kraniotis:2021qah} for the metric~\eqref{eq:metric_static}. For this reason, we introduce the Weyl tensor,  defined as the trace-free part of the Riemann tensor~\cite{Wald}
\begin{align}
C_{abcd}= R_{abcd} -\frac{1}{2} \left( g_{ac}R_{bd}-g_{ad}R_{bc}
-g_{bc}R_{ad}+g_{bd}R_{ac} \right) +\frac{R}{6} \left(
g_{ac}g_{bd}-g_{ad}g_{bc} \right)\,,
\end{align}
where $g_{ab}=\eta_{ab}$ in an orthonormal frame. A direct computation yields the independent nonvanishing components
\begin{subequations}\label{eq:weyl_re}
\begin{align}
C_{\hat{0}\hat{1}\hat{0}\hat{1}} &= \frac{1}{3}(X - 2Y - Z)\,, \\
C_{\hat{0}\hat{2}\hat{0}\hat{2}} &= C_{\hat{0}\hat{3}\hat{0}\hat{3}} 
= \frac{1}{6}(-X + 2Y + Z)\,, \\
C_{\hat{1}\hat{2}\hat{1}\hat{2}} &= C_{\hat{1}\hat{3}\hat{1}\hat{3}} 
= \frac{1}{6}(X - 2Y - Z)\,, \\
C_{\hat{2}\hat{3}\hat{2}\hat{3}} &= \frac{1}{3}(-X + 2Y + Z)\,.
\end{align}
\end{subequations}
We find that the nonvanishing ZM invariants  for the spacetime \eqref{eq:metric_static} are given by  \footnote{There is a typo in Eq.~(6c) of Ref.~\cite{Hu:2023iuw}. Using the notations of Ref.~\cite{Hu:2023iuw},  the correct expression should read $\mathcal{I}_{13}=\frac{2}{3}\mathcal{A}_3(\mathcal{A}_2^2-\mathcal{A}_1^2)^2$. }
\begin{subequations}
\begin{align}
\mathcal{I}_{1} :=& C_{abcd}C^{abcd}=\frac{4}{3} (-X+2 Y+Z)^2,\\
%\mathcal{I}_{2} :=& {}^*C_{abcd} \,C^{abcd}=0\,,\\
\mathcal{I}_{3} :=& C_{ab}{}^{cd}\,C_{cd}{}^{ef}\,C_{ef}{}^{ab}=-\frac{4}{9}  (X-2 Y-Z)^3,\\ 
%\mathcal{I}_{4} :=& -C_{ab}{}^{cd}\,{}^*C_{cd}{}^{ef}\,C_{ef}{}^{ab}=0,\\
\mathcal{I}_{5} :=& R=2(Z-X-4Y),\\
\mathcal{I}_{6} :=& R_{ab}\,R^{ab}=2(X+2Y)^2+2(2Y-Z)^2,\\
\mathcal{I}_{7} :=& R^{a}{}_{b}\,R^{b}{}_{c}\,R^{c}{}_{a}=2 \left[(Z-2 Y)^3-(X+2 Y)^3\right],\\
\mathcal{I}_{8} :=& R^{a}{}_{b}\,R^{b}{}_{c}\,R^{c}{}_{d}\,R^{d}{}_{a}=2 \left[(X+2 Y)^4+(Z-2 Y)^4\right],\\
\mathcal{I}_{9} :=& C_{abcd}R^{bc}R^{ad}=\frac{2}{3} (X+Z)^2 (X-2 Y-Z),\\
%\mathcal{I}_{10} :=& -{}^*C_{abc}{}^{d}\,R^{bc}R^{a}{}_{d}=0,\\
\mathcal{I}_{11} :=& R^{ab}R^{cd}\big(C_{eab}{}^{f}C_{fcd}{}^{e}-{}^*C_{eab}{}^{f}\,{}^*C_{fcd}{}^{e}\big)=\frac{4}{9} (X+Z)^2 (-X+2 Y+Z)^2,\\
%\mathcal{I}_{12} :=& -R^{ab}R^{cd}\big({}^*C_{eab}{}^{f}C_{fcd}{}^{o}-C_{eab}{}^{f}\,{}^*C_{fcd}{}^{o}\big)=0, \\
\mathcal{I}_{13} :=& R_{a}{}^{c} R_{c}{}^{e} R_{b}{}^{d} R_{d}{}^{f} C^{ab}{}_{ef} =-\frac{2}{3} (X+Z)^2 (X-2 Y-Z) (X+4 Y-Z)^2, \\
\mathcal{I}_{15}:=&\frac{1}{16}\,R^{ab}R^{cd}\,(C_{ea bf}C^{e}{}_{cd}{}{}{}^{f}+{}^*C_{ea bf}\, {}^*C^{e}{}_{cd}{}{}{}^{f}) = \frac{1}{36} (X+Z)^2 (-X+2 Y+Z)^2,
\\
\mathcal{I}_{16} :=& \frac{1}{32}\, R^{ab} R^{cd} 
\,C^{efgh}(C_{eabh}C_{fcdg}+ {}^*C_{eabh}  {}^*C_{fcdg}) \notag \\ =&-\frac{1}{108} (X+Z)^2 (X-2 Y-Z)^3,
%\mathcal{I}_{17} :=& \frac{1}{32}\, R^{ab} R^{cd} 
%\Big(
%  {}^*C_{efgh}\, C^{e ab h}\, C^{fcdg} 
%  + {}^*C_{efgh}\, {}^*C^{e ab h}\, {}^*C^{fcdg}  
%  \notag \\ &  - C_{efgh}\, {}^*C^{e ab h}\, C^{fcdg}  
%  + C_{efgh}\, C^{e ab h}\, {}^*C^{fcdg} 
%\Big)=0,
\end{align}
\end{subequations}
where $^*C_{abcd}$ is the dual of the Weyl tensor 
\begin{align}
{}^*C_{abcd}:=\frac{1}{2}\,E_{abef}\,C^{ef}{}_{cd},
\end{align}
with $E_{abef}$ the Levi-Civita tensor~\cite{universe6020022}. 

For completeness, we also list the remaining ZM invariants 
\begin{subequations}
\label{vanishing-ZM}
\begin{align}
\mathcal{I}_2 :=& {}^*C_{abcd} C^{abcd},    
\\
\mathcal{I}_{4} :=&C_{ab}{}^{cd}\,{}^*C_{cd}{}^{ef}C_{ef}{}^{ab},
\\
\mathcal{I}_{10} :=& {}^*C_{abc}{}^{d}R^{bc}R_{d}{}^{a},
\\
\mathcal{I}_{12} :=& 2 R^{ab} R_{cd} C_{eabf}\, {}^*C^{ecdf},
\\
\mathcal{I}_{14} :=& R_{a}{}^{c} R_{c}{}^{e} R_{b}{}^{d} R_{d}{}^{f} \, {}^*C^{ab}{}_{ef},
\\
\mathcal{I}_{17} :=&  \frac{1}{32}\, R^{ab} R^{cd} 
\,{}^*C^{efgh}(C_{eabh}C_{fcdg}+ {}^*C_{eabh}  {}^*C_{fcdg}). 
\end{align}
\end{subequations}
For the static spherically symmetric geometry \eqref{eq:metric_static},
the Weyl tensor is purely electric~\cite{Maartens:1997fg}, as is 
manifest from Eq.~\eqref{eq:weyl_re}. Hence all parity-odd dual
contractions entering the ZM scalars listed in Eq.~\eqref{vanishing-ZM} vanish identically.

Remarkably, the Kretschmann scalar can be expressed as (cf. Eq. \eqref{kretschmann-X-Y-Z})
\begin{align}
K = 4(X^2 + 4 Y^2 + Z^2), \label{Kretsch-one-funct-appendix}
\end{align} 
which constitutes a positive-definite quadratic form in $X,Y,$ and $Z$. Consequently, the finiteness of $K$ ensures that $X,Y$, and $Z$ remain finite, and hence all other algebraic curvature invariants, including the full ZM set, are finite as well. Hence, the Kretschmann scalar \emph{alone} is sufficient to fully determine the curvature regularity of the  spacetime~\eqref{eq:metric_static}. 

We can thus conclude that all solutions presented in Sec.~\ref{sec:3} have finite curvature invariants. This follows, among the assumptions underlying our procedure, from hypothesis \eqref{eq:requirerho-1}, which, jointly with Eq.~\eqref{eq:addtion_req}, guarantees  
that $X$, $Y$, and $Z$ remain finite for any value of $r$.

\section{Kretschmann scalar for generic static, spherically symmetric spacetimes}
\label{Appendix-K-ge}

In this Appendix, we provide further details on a  claim  made in Sec. \ref{Sec:2-function-setup}, namely  that the Kretschmann scalar \emph{alone} is sufficient to characterize the finiteness of all curvature scalars also for a generic static,  spherically symmetric spacetime
\begin{align}\label{eq:metric_static_generic}
{\rm d}s^2
= - B(r)\,{\rm d}t^2
+ A(r)\,{\rm d}r^2
+ r^2\,{\rm d}\Omega^2\,. 
\end{align}

Following a procedure similar to that adopted in Appendix~\ref{Appendix-0}, upon introducing the  orthonormal frame 
\begin{align}
e_{a}{}^{\mu} =
\begin{cases}
\mathrm{diag}\left[ 1/\sqrt{B},\, 1/\sqrt{A},\, 1/r,\, 1/\left(r\sin\theta\right) \right], 
\quad  \;\;\; \;\;\,\, (B > 0 ,\, A>0), & \\
\mathrm{diag}\left[ 1/\sqrt{-B},\, 1/\sqrt{-A},\, 1/r,\, 1/\left(r\sin\theta \right) \right],  \;\;\;\;\;  (B < 0,\, A<0), &
\end{cases}
\label{eq:orthonormal_frameAB}
\end{align}
one  finds the following independent nonvanishing components of the Riemann tensor: 
\begin{subequations}
\begin{align}
      R_{\hat{0}\hat{1}\hat{0}\hat{1}} &=X_1\,,\\
   R_{\hat{0}\hat{2}\hat{0}\hat{2}} &=R_{\hat{0}\hat{3}\hat{0}\hat{3}}=X_2\,,\\
   R_{\hat{1}\hat{2}\hat{1}\hat{2}} &=R_{\hat{1}\hat{3}\hat{1}\hat{3}}=X_3\,,\\
    R_{\hat{2}\hat{3}\hat{2}\hat{3}} &=X_4\,,   \end{align}  
\end{subequations}
where, for convenience, we recall  (see Eq. \eqref{eq:X14-text})
\begin{align}\label{eq:X14}
    X_1(r):=\frac{(B'/\sqrt{AB})'}{2\sqrt{AB}} \,, \quad X_2(r):=\frac{B'}{2rAB} \,, \quad 
    X_3(r):=\frac{A'}{2rA^2} \,, \quad  X_4(r):=\frac{1-1/A}{r^2}\,.
\end{align}
The remaining components again follow from the standard symmetries~\eqref{eq:Rsymmetry} of the Riemann tensor.  Consequently, any algebraic scalar  invariant can be expressed as a polynomial in  $X_1,X_2,X_3$, and $X_4$. In particular,  the Kretschmann scalar takes the compact form (cf. Eq. \eqref{eq:generic-K-positive}) 
\begin{align}
K =  4(X_1^2+2X_2^2+2X_3^2+X_4^2),\label{Kretsch-2-funct-appendix}
\end{align}
and readily reduces to the corresponding one-function expression \eqref{Kretsch-one-funct-appendix}  owing to the identifications 
\begin{equation}
  {X}_1\equiv X,
  \qquad
  {X}_2\equiv Y,
  \qquad
  {X}_3\equiv-Y,
  \qquad
  {X}_4\equiv Z.
\end{equation}

The decomposition \eqref{Kretsch-2-funct-appendix} shows explicitly that  $K$  is a positive-definite quadratic form in $X_1,X_2,X_3$, and $X_4$. Its finiteness thus guarantees the finiteness of all algebraic curvature invariants for the generic geometry \eqref{eq:metric_static_generic}, mirroring the one-function setup studied in Appendix \ref{Appendix-0}. Our analysis agrees with the observation made in Refs.~\cite{Bronnikov:2012wsj,Lobo2020f} 
\footnote{After completing the calculation presented in this appendix, we became aware that a decomposition of the Kretschmann scalar into a positive semi-definite sum of squares had already appeared in the literature; see, e.g., Refs.~\cite{Bronnikov:2012wsj,Gkigkitzis:2014cla}. With this decomposition, it follows that, in static and spherically symmetric settings, the finiteness of the Kretschmann scalar is sufficient to ensure that all algebraic curvature invariants remain finite~\cite{Bronnikov:2012wsj,Lobo2020f}.}.

\bibliographystyle{apsrev4-1}
\bibliography{references}{}

\end{document}